\documentclass[12pt]{article}
\usepackage{amsmath, amssymb}
\usepackage{mathrsfs}
\usepackage[dvips]{graphicx}
\usepackage[nosort]{cite}

\def\del{\partial }  
\def\no {\nonumber}

\def\B {\mathcal{B}}

\def\P {\mathcal{P}}

\def\lB {\mathcal{B}_{\mathrm{lin}}}

\begin{document}

\baselineskip=17pt

\begin{titlepage}
\rightline{\tt arXiv:1207.6220}
\rightline{\tt UT-Komaba/12-7}
\rightline{\tt RUP-12-7}
\begin{center}
\vskip 2.5cm
{\Large \bf {Constraints on a class of classical solutions}}\\
\vskip 0.4cm
{\Large \bf {in open string field theory}}
\vskip 1.0cm
{\large {Toru Masuda$^1$, Toshifumi Noumi$^1$ and Daisuke Takahashi$^2$}}
\vskip 1.0cm
$^1${\it {Institute of Physics, The University of Tokyo}}\\
{\it {Komaba, Meguro-ku, Tokyo 153-8902, Japan}}\\
\vskip 0.4cm
$^2${\it Department of Physics, Rikkyo University, Tokyo 171-8501, Japan}\\
\vskip 0.4cm
masudatoru@gmail.com,
tnoumi@hep1.c.u-tokyo.ac.jp,
d-takahashi@rikkyo.ac.jp

\vskip 3.0cm

{\bf Abstract}
\end{center}
\noindent
We calculate boundary states for general string fields
in the $KBc$ subalgebra under some regularity conditions
based on the construction by Kiermaier, Okawa, and Zwiebach. 
The resulting boundary states are always proportional to that
for the perturbative vacuum~$|B\rangle$.
In this framework,
the equation of motion implies that
boundary states are independent of the 
auxiliary 
parameter $s$
associated with the length of the boundary. 
By requiring the $s$-independence,
we show that the boundary states
for classical solutions in our class
are restricted to
$\pm|B\rangle$ and $0$. 
In particular, there exist no string fields which reproduce boundary states for multiple D-brane backgrounds.
While we know that the boundary states
$|B\rangle$ and $0$ are reproduced by
solutions for the perturbative vacuum  and the tachyon vacuum, respectively,
no solutions
reproducing $-|B\rangle$ 
have been constructed.
In this paper
we also
propose
a candidate 
for such a 
solution,
which may describe the ghost D-brane.

\end{titlepage}

\newpage

\allowdisplaybreaks

\setcounter{tocdepth}{2}
{\small
\tableofcontents 
}

\section{Introduction and summary}
\setcounter{equation}{0}
String field theory is a field theoretical approach to
non-perturbative aspects of string theory.
Its classical solutions describe
consistent backgrounds of the string,
and
it can potentially give a framework to explore various string vacua.

\medskip
In the case of open string field theory~\cite{Witten:1985cc},
considerable understanding of the landscape
has been obtained
especially since
Schnabl constructed an analytic solution
for the tachyon vacuum~\cite{Schnabl:2005gv},
where unstable D-branes have disappeared
and there are no physical excitations of the open string.
It is notable that
classical open string field theory can describe
the decay process of D-branes~\cite{Schnabl:2007az,Kiermaier:2007ba,Erler:2007rh,
Okawa:2007ri,Okawa:2007it,Ellwood:2007xr,Kwon:2008ap,
Hellerman:2008wp,Kishimoto:2008zj,Barnaby:2008pt,
Beaujean:2009rb,Kiermaier:2010cf,Noumi:2011kn}
as well as the tachyon vacuum~\cite{Schnabl:2005gv,Ellwood:2006ba,Erler:2009uj}
without explicit closed string degrees of freedom.
Since closed strings are considered to be emitted in the D-brane decay process,
classical open string theory is expected to have
some information about the closed string.
By further investigating classical solutions in open string field theory,
we would like to explore the scope of open string
theory.

\medskip 
Schnabl's original solution was constructed from a class of
wedge-based states.\footnote{
We denote wedge states~\cite{Rastelli:2000iu,Schnabl:2002gg} with operator insertions by wedge-based states.
}
When we write the wedge state $W_\alpha$ as $W_\alpha = e^{\alpha K}$,\footnote{
Products of string fields in this paper are defined
using Witten's star product~\cite{Witten:1985cc}.
}
the solution can be written in terms of the states $K$, $B$, and $c$.
These states are associated with the energy-momentum tensor,
the $b$ ghost, and the $c$ ghost, respectively.\footnote{
A precise definition of these states are given in section~\ref{section_setup}.}
They satisfy
the following simple algebraic relations
called the $KBc$ subalgebra~\cite{Okawa:2006vm}:
\begin{equation}
\label{KBc_sect1}
B^2=c^2=0\,,\quad \{B,c\}=1\,,\quad
QB=K\,,\quad
Qc=cKc\,,\quad
[K,B]=0\,,
\end{equation}
where $Q$ is the BRST operator of the open bosonic string.
These algebraic relations and their extension have been
the starting point to construct analytic solutions in open string field theory~\cite{Schnabl:2007az,Kiermaier:2007ba,Erler:2007rh,Okawa:2007ri,Okawa:2007it,
Kiermaier:2010cf,Noumi:2011kn,Erler:2009uj,
Rastelli:2006ap
,Okawa:2006sn,Erler:2006hw,Erler:2006ww,
Fuchs:2007yy,Fuchs:2007gw,Kiermaier:2007vu,
Erler:2007xt,Kiermaier:2007ki,Aref'eva:2008ad,
Aref'eva:2009ac,Ellwood:2009zf,Arroyo:2010fq,
Erler:2010pr,Bonora:2010hi,Murata:2011ex,
Bonora:2011ri,Erler:2011tc,Murata:2011ep}
. An important property of the $KBc$ subalgebra is that
we can define the states
without specifying the D-brane configuration at the perturbative vacuum.
Furthermore,
the algebraic relations~(\ref{KBc_sect1})
follow only from the operator product expansions
of the energy momentum tensor,
the $b$ ghost, and the $c$ ghost.
In this sense,
the $KBc$ subalgebra is in the universal sector of
open bosonic string field theory~\cite{Sen:1999xm},
which is expected to contain
classical solutions such as those for the tachyon vacuum
and multiple D-branes.

\medskip 
In~\cite{Okawa:2006vm},
Okawa proposed a class of formal solutions
in the $KBc$ subalgebra
as a generalization of Schnabl's solution:
\begin{equation}
\label{Okawa's_solution}
\Psi=F(K)c\frac{KB}{1-F(K)^2}cF(K)\,,
\end{equation}
where $F(K)$ is a function of $K$.
This form of string fields can be formally written
in the pure-gauge form,
\begin{equation}
\Psi=U^{-1}QU\quad
{\rm with}
\quad
U=1-F(K)cBF(K)\,,
\end{equation}
and therefore,
they are expected to satisfy the equation of motion
of open bosonic string field theory:
\begin{equation}
\label{eom}
Q\Psi+\Psi^2=0\,.
\end{equation}
Recently in~\cite{Murata:2011ex,Murata:2011ep},
Murata and Schnabl systematically studied
this class of 
solutions
and
made an interesting proposal for
multiple D-brane solutions.
They evaluated energy and
a kind of
gauge-invariant observables~\cite{Hashimoto:2001sm,Gaiotto:2001ji,Ellwood:2008jh}
and
found that those
for $n$ D-branes are reproduced
by choosing the function $F(K)$ such that
\begin{equation}
\lim_{z\to0}z\frac{\left(F(z)^2\right)^\prime}{1-F(z)^2}=1-n\label{MS_result}\,.
\end{equation}
In fact,
the proposed 
solutions are singular
and 
require
some regularization.
In order to make the solutions fully acceptable,
one has to regularize them
so that
they reproduce
physical quantities such as
energy and the gauge-invariant observables
without ambiguity
as well as they respect the equation of motion.
However,
despite some efforts~\cite{Murata:2011ep,Hata:2011ke,Masuda},
no regularization method consistent with
all the above requirements is known.

\medskip
In~\cite{Takahashi:2011wk},
one of the authors evaluated the boundary states
for string fields in the form of~(\ref{Okawa's_solution})
based on the construction by Kiermaier, Okawa, and Zwiebach\cite{Kiermaier:2008qu}.\footnote{
Recently in~\cite{Kudrna:2012re},
another interesting approach to construct
boundary states from open string classical solutions
was proposed.}
It was found that 
the proposed multiple D-brane solutions
do not reproduce the expected boundary states.
It was also found that the boundary states non-trivially depend
on the parameter $s$
associated with the length of the boundary.
This is a serious problem
because the non-trivial $s$-dependence
of the boundary states
indicates violation of
the equation of motion~\cite{Kiermaier:2008qu}.
These results suggest a difficulty
in the construction of multiple D-brane solutions in the form of~(\ref{Okawa's_solution})
without additional terms.

\medskip
The purpose of this paper is to
further develop the argument in~\cite{Takahashi:2011wk}
and
clarify which class of classical solutions can reproduce 
physically desired boundary states. 
By requiring the $s$-independence of the boundary state
as a necessary condition
to satisfy the equation of motion,
we investigate possible
solutions 
in the $KBc$ subalgebra.
Extending the calculation in~\cite{Takahashi:2011wk},
we evaluate the boundary states
for general string fields
in the $KBc$ subalgebra
given by
\begin{equation}
\label{general_Psi_1}
\Psi=\sum_iF_i(K) cB G_i(K) c H_i(K)
\end{equation}
under some regularity conditions introduced in section~\ref{subsection_regularity}.
The obtained boundary state\footnote{
To be precise,
the closed string state $|B_*(\Psi)\rangle$ constructed by Kiermaier, Okawa, and Zwiebach corresponds to 
the boundary state up to BRST-exact terms. 
Moreover,
it does not satisfy some 
properties of boundary states
unless $\Psi$ satisfies the equation of motion.
However, in this section,
we call it
the boundary state for~simplicity.
}
for the string field~(\ref{general_Psi_1}) is given by
\begin{equation}
\label{obtained_bdst}
|B_*(\Psi)\rangle
=\frac{e^{(x+1)s} - e^{y s}}{e^s - 1}|B\rangle\;,
\end{equation}
where $|B\rangle$ is the boundary state for the perturbative vacuum
and the parameter $s$ is associated with the length of the boundary.
The $c$-numbers $x$ and $y$ are
given by
\begin{align}
\label{alpha_1}
x&=\sum_{i}G_i(0)\left(\frac{1}{2}F_i(0)H_i(0)+F_i^\prime(0) H_i(0)\right)\,,\\
\label{beta_1}
y&=\sum_{i}G_i(0)\left(\frac{1}{2}F_i(0)H_i(0)-F_i(0) H_i^\prime(0)\right)\,.
\end{align}
Note that we do not use 
the equation of motion
to derive the formulae \eqref{obtained_bdst}-\eqref{beta_1}. 
As we mentioned earlier,
the boundary state $|B_\ast(\Psi)\rangle$
does not depend on the parameter $s$
when the string field~(\ref{general_Psi_1}) satisfies
the equation of motion~(\ref{eom}).
It is not difficult to see that the state~(\ref{obtained_bdst})
is $s$-independent
only in the following three cases:
\begin{equation}
\nonumber
\begin{array}{lcll}\bullet\quad | B_\ast(\Psi)\rangle\,\,=\,\,| B\rangle & \quad{\rm for}\quad & x=y=0\,,\\[2mm]
\bullet\quad | B_\ast(\Psi)\rangle\,\,=\,\,0 & \quad{\rm for}\quad & x=y-1={\rm arbitrary}  \,,\\[2mm]
\bullet\quad | B_\ast(\Psi)\rangle\,\,=\,\,-| B\rangle & \quad{\rm for}\quad & x=-1\,,\,\,y=1\,.
\end{array}
\end{equation}
We know that
the boundary states for the first two cases
are realized by solutions for the perturbative vacuum
and the tachyon vacuum, respectively~\cite{Takahashi:2011wk,Kiermaier:2008qu}.
However,
we do not know what kind of solutions
in open string field theory
reproduce that for the third case.
In~\cite{Okuda:2006fb},
Okuda and Takayanagi
introduced so-called ghost D-branes.
They have negative tension
and the corresponding boundary state
is that for the perturbative vacuum with additional minus sign. 
Following the paper~\cite{Okuda:2006fb},
we call solutions reproducing such a boundary state ghost brane solutions.
Our result suggests that
all the classical solutions in the $KBc$ subalgebra can be classified into above three cases.
In particular,
there seems to exist no solution reproducing the boundary states
for multiple D-brane backgrounds in the $KBc$ subalgebra.\footnote{
We should mention that
our formulae~\eqref{obtained_bdst}-\eqref{beta_1}
is derived using the prescription
developed in~\cite{Kiermaier:2008jy}
based on the Schnabl gauge propagator.
Although this prescription is well-established for solutions
constructed from wedge-based states of finite width,
it is not clear whether it is applicable for those containing
wedge-based states of infinite width.
}
This is the first main result of this paper.

\medskip
In the above discussion,
we classified
possible boundary states
in the $KBc$ subalgebra
by requiring $s$-independence as a necessary condition.
In order to investigate concrete expression
of classical solutions reproducing those boundary states,
we consider two classes of formal solutions.
We first consider the following class: 

\begin{align}
\label{formally_pure_gauge_1}
\Psi=U^{-1}QU=\sum_{i}I_icKBcJ_i+\sum_{i}I_icJ_i\frac{KB}{1-\sum_{j}I_jJ_j}\sum_{k}I_kcJ_k\,,
\end{align}
where 
the gauge parameter $U$ is given by
\begin{equation}
U=1-\sum_{i}I_icB\,J_i\,.
\end{equation}
Note that when the indices $i$ and $j$ run over only one value,
these formal solutions reproduce
Okawa's formal solutions~(\ref{Okawa's_solution}).
Here,
$I_i=I_i(K)$ and $J_i=J_i(K)$,
and
we sometimes omit the explicit indication
of the argument~$K$
in the rest of this paper
for simplicity.
Applying our general formulae \eqref{obtained_bdst}-\eqref{beta_1}
of the boundary states
to the string field~(\ref{formally_pure_gauge_1}),
we obtain that
\begin{equation}
\nonumber
\begin{array}{lcll}\bullet\quad | B_\ast(\Psi)\rangle\,\,=\,\,| B\rangle & \quad{\rm for}\quad & \sum_j I_j(0) J_j(0)\neq1\,,\\[2mm]
\bullet\quad | B_\ast(\Psi)\rangle\,\,=\,\,0 & \quad{\rm for}\quad & \sum_j I_j(0) J_j(0)=1\quad {\rm and}\quad
\sum_j\left(I_j^\prime(0) J_j(0) + I_j(0) J_j^\prime(0)\right)\neq 0  \,.\end{array}
\end{equation}
The formal solution 
(\ref{formally_pure_gauge_1}) does not satisfy
our regularity conditions
when the functions $I_i(K)$ and $J_i(K)$ satisfy $\sum_j I_j(0) J_j(0)=1$ and $\sum_j\left(I_j^\prime(0) J_j(0) + I_j(0) J_j^\prime(0)\right)=0$.
In other words, this class of string fields
satisfying our regularity conditions
do not contain the ghost brane solution,
and
they reproduce the boundary states
for the perturbative and the tachyon vacuum only.
Note that our results seems to consistent with Erler's classification \cite{Erler:2006ww} of Okawa-type solutions \eqref{Okawa's_solution}
with respect to the energy calculation.

\medskip
We next consider another class of formal solutions:
\begin{equation}\label{another class of solutions}
\Psi=Fc\frac{KB}{1-F^2}cF+Hc\frac{KB}{1-H^2}cH
\quad{\rm with}\quad FH=1\,,
\end{equation}
which cannot be written in the form~(\ref{formally_pure_gauge_1}).
Applying our general formulae \eqref{obtained_bdst}-\eqref{beta_1},
we find that
\begin{equation}
\label{another_intro}
\begin{array}{lcll}\bullet\quad | B_\ast(\Psi)\rangle\,\,=\,\,| B\rangle & \quad{\rm for}\quad & F(0)\neq1\,,\\[2mm]
\bullet\quad | B_\ast(\Psi)\rangle\,\,=\,\,-| B\rangle & \quad{\rm for}\quad & F(0)=1 \quad {\rm and}\quad
F^{\prime}(0)\neq 0  \,.\end{array}
\end{equation}
The formal solutions
(\ref{another class of solutions}) does not satisfy
our regularity conditions
when the function $F(K)$ satisfies $F(0)=1$ and $F^\prime(0)=0$.
The first case reproduces the boundary state for the perturbative vacuum
although
it seems not to be gauge-equivalent to the perturbative vacuum
as we mention later.\footnote{
We showed that
only the boundary states
for the perturbative vacuum, the tachyon vacuum, and the ghost D-brane
can be reproduced by classical solutions satisfying our regularity conditions.
However,
as the first case in~(\ref{another_intro}) shows,
it does not necessarily mean that solutions in our class are restricted
to those for the three backgrounds.}
The second case is nothing but the ghost brane solution.
We propose a concrete candidate for the ghost brane solution in the following form:
\begin{equation}
\label{ghost_brane_1}
\Psi_{\it{ghost}}
=\sqrt{\frac{1-pK}{1-qK}}c\frac{(1-qK)}{p-q}Bc\sqrt{\frac{1-pK}{1-qK}}+
\sqrt{\frac{1-qK}{1-pK}}c\frac{(1-pK)}{q-p}Bc\sqrt{\frac{1-qK}{1-pK}}\,,
\end{equation}
where $p$ and $q$ are distinct positive constants.
The state ${\sqrt{(1-pK)/(1-qK)}}$\, 
is defined by the superposition of wedge states:
\begin{align}
\sqrt{\frac{1-pK}{1-qK}}
&=\int_0^\infty dt_1\int_0^\infty dt_2 \, \frac{e^{-t_1}}{\sqrt{\pi t_1}}\frac{e^{-t_2}}{\sqrt{\pi t_2}} \, e^{(qt_1+pt_2)K}(1-pK)\,.
\end{align}
We carefully
show that
the ghost brane solution (\ref{ghost_brane_1}) satisfies the equation of motion and it has definite energy density
of minus two times the D-brane tension.\footnote{
The ghost D-brane discussed in~\cite{Okuda:2006fb}
has the same energy density
and reproduces the same boundary state
as our ghost brane solution.}
Although its physical interpretation is still obscure,
these results seem to be consistent with 
each other; 
we have no definite reason to rule out them.
This is the second main result of this paper.

\medskip
The organization of this paper is as follows.
In section~\ref{section_setup} we introduce the $KBc$ subalgebra
and our regularity conditions on string fields.
In section~\ref{boundarystate_sec}
we evaluate the boundary states
for general string fields in the $KBc$ subalgebra
under the regularity conditions introduced in section~\ref{subsection_regularity}.
After reviewing the construction by Kiermaier, Okawa, and Zwiebach~\cite{Kiermaier:2008qu},
we derive the general formulae \eqref{obtained_bdst}-\eqref{beta_1}. 
By requiring $s$-independence as a necessary condition
to satisfy the equation of motion,
we show that possible solutions are only those reproducing
the boundary states for the perturbative vacuum,
the tachyon vacuum,
and the ghost brane.
In section~\ref{pure_sec}
we consider the formal solutions~\eqref{formally_pure_gauge_1}.
We show that
this class of solutions include only those reproducing
the boundary states for the tachyon vacuum and the perturbative vacuum.
In section~\ref{ghost_sec} we consider another class of formal solutions \eqref{another class of solutions},
and we propose a candidate for the ghost brane solution.  
Section~\ref{discussion_sec} is devoted to discussion. 
\section{Setup}
\label{section_setup}

\setcounter{equation}{0}
\subsection{$KBc$ subalgebra}
The wedge state $W_\alpha$ with $\alpha \ge 0$ is defined by
its BPZ inner product $\langle \, \varphi, W_\alpha \, \rangle$ as follows:
\begin{equation}
\label{wedge-state}
\langle \, \varphi,W_\alpha \, \rangle
=\left\langle \, \mathcal{F}\circ \varphi(0) \, \right\rangle_{C_{\alpha+1}}.
\end{equation}
Here and in what follows $\varphi$ denotes
a generic state in the Fock space
and its corresponding operator is $\varphi(\xi)$ in the state-operator mapping.
We denote the conformal transformation of $\varphi (\xi)$
under the map $\mathcal{F}(\xi)$ by $\mathcal{F} \circ \varphi (\xi)$, where
\begin{equation}
\mathcal{F}(\xi) = \frac{2}{\pi} \, \arctan \xi \,.
\end{equation}
The coordinate $z$ related through $z = \mathcal{F}(\xi)$
to the coordinate $\xi$
on the upper half-plane used in the standard state-operator mapping
is called the sliver frame.
The correlation function is evaluated on the surface $C_{\alpha+1}$,
which is the semi-infinite strip obtained from the upper half-plane of $z$
by the identification $z \sim z+\alpha+1$.
We usually use the region $-1/2 \le \Re z \le 1/2+\alpha$ for $C_{\alpha+1}$.

\medskip
Just as the line integral $L_0$ of the energy-momentum tensor
generates a surface $e^{-t L_0}$ in the standard open string strip coordinate,
the wedge state $W_\alpha$ can be thought of as being generated
by a line integral of the energy-momentum tensor in the sliver frame.
We denote the wedge state $W_0$ of zero width
with an insertion of the line integral by $K$
and we write the wedge state $W_\alpha$~as
\begin{equation}
W_\alpha = e^{\alpha K} \,.
\end{equation}
An explicit definition of the state $K$ is given by
\begin{equation}
\left\langle\varphi,K\right\rangle
=\biggl\langle \, \mathcal{F}\circ \varphi(0)\int_{\frac{1}{2}+i\infty}^{\frac{1}{2}-i\infty}\frac{dz}{2\pi i} \, T(z) \, \biggr\rangle_{C_{1}},
\end{equation}
where $T(z)$ is the energy-momentum tensor
and we use the doubling trick.
Note that the line integral is from a boundary to the open string mid-point
before using the doubling trick, while the line integral $L_0$ is from a boundary
to the other boundary.

\medskip
Just as the line integral $L_0$ is the BRST transformation
of the line integral $b_0$ of the $b$ ghost,
the line integral that generates the wedge state
is the BRST transformation of the same line integral
with the energy-momentum tensor replaced by the $b$ ghost.
Correspondingly, we define the state $B$ by
\begin{equation}
\langle \, \varphi,B \, \rangle
=\biggl\langle \, \mathcal{F}\circ \varphi(0)\int_{\frac{1}{2}+i\infty}^{\frac{1}{2}-i\infty}\frac{dz}{2\pi i} \, b(z) \, \biggr\rangle_{C_{1}}.
\end{equation}
By construction, the state $K$ is the BRST transformation of $B$.
Another important property of the state $B$ is that $B^2 = 0$.

\medskip
We also define the state $c$ of ghost number $1$ by
a state based on the wedge state $W_0$
with a local insertion of $c(t)$ on the boundary.
More explicitly,
it is given by
\begin{equation}
\langle \, \varphi,c \, \rangle
=\biggl\langle \, \mathcal{F}\circ \varphi(0)\,c\,(\tfrac{1}{2}) \, \biggr\rangle_{C_{1}}.
\end{equation}
The state $c$ satisfies~\cite{Okawa:2006vm}\footnote{
Our definition of the states
can be related to that in~\cite{Okawa:2006vm}
as
$K_{\rm here} = (\pi/2) \, K_{\rm there}$,
 $B_{\rm here} = (\pi/2) \, B_{\rm there}$,
 and
 $c_{\rm here} = (2/\pi) \, c_{\rm there}$.}

\begin{equation}
Qc=cKc\,,\quad
c^2=0\,,\quad\{B,c\}=1\,,
\end{equation}
which follow from
the BRST transformation of the $c$ ghost,
$Q\cdot c(t)=c\partial c(t)$,
and
the operator product expansions
of the energy-momentum tensor,
the $b$-ghost,
and the $c$-ghost.

\medskip
To summarize,
the states $K$, $B$, and $c$
satisfy the following algebraic relations called the $KBc$ subalgebra:
\begin{equation}
\label{KBc}
\begin{split}
B^2=c^2=0\,,\quad \{B,c\}=1\,,\quad
QB=K\,,\quad
QK=0\,,\quad
Qc=cKc\,,\quad
[K,B]=0\,.
\end{split}
\end{equation}
As we mentioned in the introduction,
we can define the states $K$, $B$, and $c$
without specifying the
D-brane configuration at the perturbative vacuum,
and they always satisfy the algebraic relations~(\ref{KBc}).

\subsection{Regularity conditions on string fields}
\label{subsection_regularity}
In the calculation of boundary states,
we require some regularity conditions
on string fields.
In this subsection
we introduce those conditions
and clarify our setup.

\paragraph{-- Regularity conditions}
A state $F(K)$ defined by a superposition of wedge states
is characterized by the following function $f(t)$:
\begin{equation}
F(K)=\int_0^\infty dt\,f(t)e^{tK}\,.
\end{equation}
In the calculation of boundary states,
we often encounter the expressions $F(0)=\int_0^\infty dt\,f(t)$
and $F^\prime(0)=\int_0^\infty dt\,f(t)\,t$.
We define two kinds of regularity conditions I and II
by
\begin{eqnarray}
\label{regularity}
\text{condition I}&:&\text{$\int_0^\infty dt\,f(t)$ is absolutely convergent}\,,\\
\label{differentiability}
\text{condition II}&:&\text{$\int_0^\infty dt\,f(t)\,t$ is absolutely convergent}\,.
\end{eqnarray}
These conditions guarantee the finiteness of $F(0)$
and $F^\prime(0)$, respectively.

\paragraph{-- Examples}
Typical examples satisfying both of the two conditions are given by
\begin{equation}
\begin{split}
e^{sK}=\int_0^\infty dt\,\delta(t-s)e^{tK}\,,
&\quad\quad
K=-\lim_{\epsilon\to0}\int_0^\infty dt\,\delta^\prime(t-\epsilon)e^{tK}\,,\\[2mm]
\frac{1}{1-K}=\int_0^\infty dt\,e^{-t}e^{tK}\,,
&\quad\quad
\frac{1}{\sqrt{1-K}}= \int_0^\infty dt \, \frac{e^{-t}}{\sqrt{\pi t}} \, e^{t K} \,.
\end{split}
\end{equation}
When we define the sliver state $\Omega^\infty$ 
and the state $\sqrt{-K}$ 
by
\begin{align}
\Omega^\infty&=\lim_{\Lambda\to\infty}e^{\Lambda K}=\lim_{\Lambda\to\infty}\int_0^\infty dt\,\delta(t-\Lambda)e^{tK}\,,\\[2mm]
\sqrt{-K}&=-\lim_{\epsilon\to0}\int_\epsilon^\infty dt\frac{1}{\sqrt{\pi t}}Ke^{tK}
=\lim_{\epsilon\to0}\int_0^\infty dt\left(\frac{\delta(t-\epsilon)}{\sqrt{\pi t}}-\frac{1}{2\sqrt{\pi}}\theta(t-\epsilon)\,t^{-3/2}\right)e^{tK}\,,
\end{align}
they satisfy the condition~I,
but they do not satisfy the condition II.
When we define the state $1/K$ by $1/K=\lim_{\epsilon\to0}\frac{1}{K-\epsilon}$,
it satisfies none of the two conditions:
\begin{equation}
\frac{1}{K}
=-\lim_{\epsilon\to0}\int_0^\infty dt e^{-\epsilon t}e^{tK}\,.
\end{equation}

\paragraph{-- Regular string fields of ghost number one}
Using the algebraic relations~(\ref{KBc}),
any string field $\Psi$ of ghost number $1$
in the $KBc$ subalgebra can be written as\footnote{
To be rigorous,
the use of the algebraic relation $\{B,c\}=1$ can affect regularity of
string fields in general.
For example,
we can rewrite
a string field $\Psi$ in the form of
$\Psi=e^{K/2}cB\frac{1}{K}cBKce^{K/2}$
as
$\Psi=e^{K/2}cBce^{K/2}$,
formally using the relation $\{B,c\}=1$ and $\frac{1}{K}K=1$.
Here, the regularity of $\Psi$ seems
to be changed by the use of $\{B,c\}=1$.
In this paper,
we assume that functions of $K$ in the string fields
are defined by superpositions of wedge states
and they
satisfy the regularity condition I.
Under this assumption,
we can use the relation $\{B,c\}=1$ without subtlety
and
any string field of ghost number $1$
can be written in the form of (\ref{general_KBc_1}).
}
\begin{equation}
\label{general_KBc_1}
\Psi=\sum_iF_i(K) cB G_i(K) c H_i(K)\,,
\end{equation}
where $F_i(K)$, $G_i(K)$, and $H_i(K)$ are functions of $K$.
The summation symbol $\sum_i$
stands for sums and integrals over the labels of the string fields.
In this paper,
we consider string fields (\ref{general_KBc_1}) 
with the states $F_i(K)$, $G_i(K)$, and $H_i(K)$
being superpositions of wedge states:
\begin{equation}
F_i(K)=\int_0^\infty dt\,f_i(t)e^{tK}\,,
\quad
G_i(K)=\int_0^\infty dt\,g_i(t)e^{tK}\,,
\quad
H_i(K)=\int_0^\infty dt\,h_i(t)e^{tK}\,.
\end{equation}
We also assume that
$F_i(K)$ and $H_i(K)$ satisfy both of the conditions
I and II,
and $G_i(K)$ satisfies the condition I.

These conditions are all we need to assume when we calculate boundary states. 
With these conditions, we can safely calculate the boundary state for $\Psi$.
Note that our regularity conditions do not exclude identity-based string fields, which are usually thought to be singular with respect to energy calculation. 
Accordingly, our calculation of boundary states can be also applicable to identity-based solutions.
In the next section
we evaluate the boundary state for~(\ref{general_KBc_1})
under these assumptions.

\section{Boundary states in $KBc$ subalgebra}
\label{boundarystate_sec}
\setcounter{equation}{0}
In this section
we discuss possible boundary states
which can be reproduced from general string fields in the
$KBc$ subalgebra
under the regularity conditions introduced in section~\ref{subsection_regularity}.
After reviewing the construction
of boundary states
in~\cite{Kiermaier:2008qu},
we derive the general formulae~\eqref{obtained_bdst}-\eqref{beta_1}.
By requiring the $s$-independence as a necessary condition
to satisfy the equation of motion,
we show that possible classical solutions
are only those
reproducing the boundary states 
for the perturbative vacuum,
the tachyon vacuum, and the ghost brane.

\subsection{Review}
In this subsection
we briefly
review the construction of boundary states proposed by Kiermaier, Okawa, and Zwiebach (KOZ)~\cite{Kiermaier:2008qu}. 
Since it is slightly complicated,
we start from the basic strategy
of the construction.

	\begin{figure}[t]
	  		\begin{center}
   				\includegraphics[width=150mm]{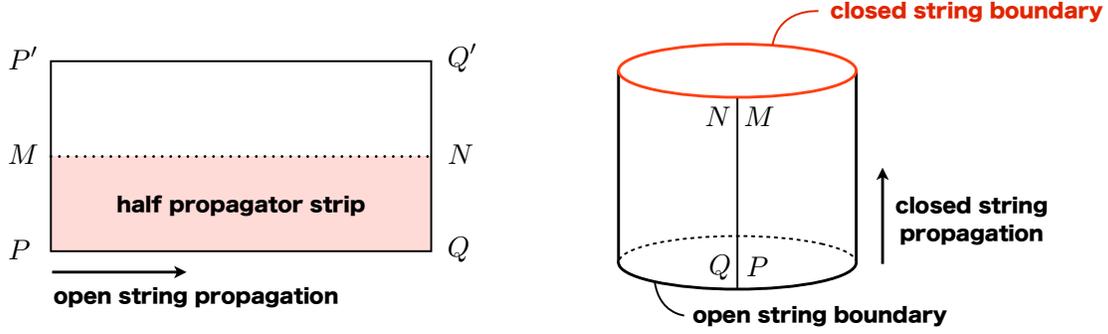}
  			\end{center}
			\vspace{-5mm}
  				\caption{
				The rectangular $PQQ^\prime P^\prime $
				is the open string world-sheet strip~$e^{-s\mathcal L}$,
				and
				the dotted line~$MN$ is the midpoint propagation
				of the open string.
				The rectangular~$PQNM$ describes the half propagator strip.
		Identifying $MP$ and $NQ$,
		we obtain a cylinder in the right figure.		
				}
  		 	\label{fig:half_propagator}
	\end{figure}

\paragraph{-- Basic strategy}
We first construct
the boundary state $|B\rangle$
for the boundary conformal field theory (BCFT)
corresponding to the perturbative vacuum.
Consider a open string world-sheet strip $e^{-s\mathcal L}$
associated with the propagator
of the open string.
In the linear $b$ gauge,
the generator $\mathcal L$ of the world-sheet strip
is given by $\mathcal{L} = \{Q, \lB\}$,
where $\lB$ denotes the ghost number $-1$ operator,
which determines the gauge condition:
\begin{equation}
\label{linear_b}
\lB\Psi=0
\quad
{\rm with}
\quad
\lB=\oint\frac{d\xi}{2\pi i}v(\xi)b(\xi)\,.
\end{equation}
We then cut the strip along
the trajectory of the midpoint propagation of the open string.
We call one of the resulting pieces {\it a half propagator strip}
(see Figure~\ref{fig:half_propagator}).  
By identifying the two edges $MP$ and $NQ$,
which represent the initial and final half-string states,
we obtain a cylinder.
One of two boundaries
of the cylinder
is the open string boundary,
where the boundary conditions are defined by
the original BCFT,
and
the other is the trajectory of the midpoint propagation,
which we call a closed string boundary.
Path integrals over the cylinder define a closed string state at
the closed string boundary. 
Then, this closed string state
reproduces the
boundary state $|B\rangle $ of the original BCFT
after an appropriate exponential action of $L_0 + \tilde L_0$,
which generates the closed string propagation. 
Note that
we can arbitrarily choose a gauge condition
and
a parameter $s$ representing 
the open string propagation
in the construction. 

\medskip
Next,
let us consider the boundary state $|B_\ast\rangle$
for the BCFT$_\ast$
associated with a classical solution $\Psi$.
The open string propagation around the background $\Psi$
is generated by $\mathcal L_\ast=\{Q_\ast , \lB\}$,
where $Q_\ast$ is the kinetic operator around the background $\Psi$.
Then, it is expected that
the boundary state $|B_*\rangle$ for BCFT$_\ast$
can be obtained by replacing $\mathcal L$
with $\mathcal L_\ast$ in the previous construction.
In~\cite{Kiermaier:2008qu},
it was discussed that
the closed string state $|B_\ast(\Psi)\rangle$
constructed in this way
coincides with the boundary state $|B_\ast\rangle$
up to some BRST-exact terms.
That's the basic strategy of the KOZ construction.

\paragraph{-- Properties} 
Before going to the concrete description of the construction,
we mention some properties of the closed string state $|B_*(\Psi) \rangle$ of the KOZ construction.
In~\cite{Kiermaier:2008qu},
it was shown that
$|B_*(\Psi) \rangle$ satisfies
the following three properties,
which can be considered as consistency requirements
for its interpretation as a boundary state:
\begin{align} 
\mathcal{Q}|B_*(\Psi) \rangle=0\,,
\quad
(b_0-\tilde{b}_0)|B_*(\Psi) \rangle=0\,,
\quad
(L_0-\tilde{L}_0)|B_*(\Psi) \rangle=0\,,
\end{align}
where $\mathcal{Q}$ is the BRST operator
of the closed bosonic string.
When two classical solutions are gauge-equivalent,
corresponding 
closed string states $|B_*(\Psi) \rangle$
are expected to be physically equivalent.
Indeed,
$|B_*(\Psi) \rangle$ is invariant under the gauge transformation $\delta_\chi$
of the classical solution $\Psi$
up to 
$\mathcal{Q}$-exact terms:
\begin{align}
\delta_\chi |B_*(\Psi) \rangle=
\mathcal{Q}\text{-exact}\,.
\end{align}
In the construction,
we can choose any regular gauge condition of the open string propagator
and the parameter $s$ associated with the propagation length.
It is expected that
physical properties of the state $|B_*(\Psi) \rangle$
do not depend on these choices.
Indeed,
it is invariant
up to $\mathcal{Q}$-exact terms under 
a variation $\delta \lB$ of the gauge condition
and 
a variation of 
the parameter~$s$:
	\begin{align}
                \delta_{\lB} |B_*(\Psi)\rangle = 
                \mathcal{Q}\text{-exact}\,,
                \quad
                \del_s |B_*(\Psi)\rangle = 
                \mathcal{Q}\text{-exact}\,.
\label{s_vari_bdy}
        \end{align}
We emphasize that
the equation of motion for $\Psi$
was used to derive all the above properties.

\medskip
We also note that
the calculation of $|B_*(\Psi) \rangle$ in the limit $s\to 0$
can be directly related to
that of the gauge invariant observables~\cite{Hashimoto:2001sm,Gaiotto:2001ji},  
which are conjectured~\cite{Ellwood:2008jh} to
represent the difference between
on-shell closed string one-point functions on the disk
for BCFT and BCFT$_\ast$.

\subsubsection{Construction}
\begin{figure}[t]
	  		\begin{center}
   				\includegraphics[width=70mm]{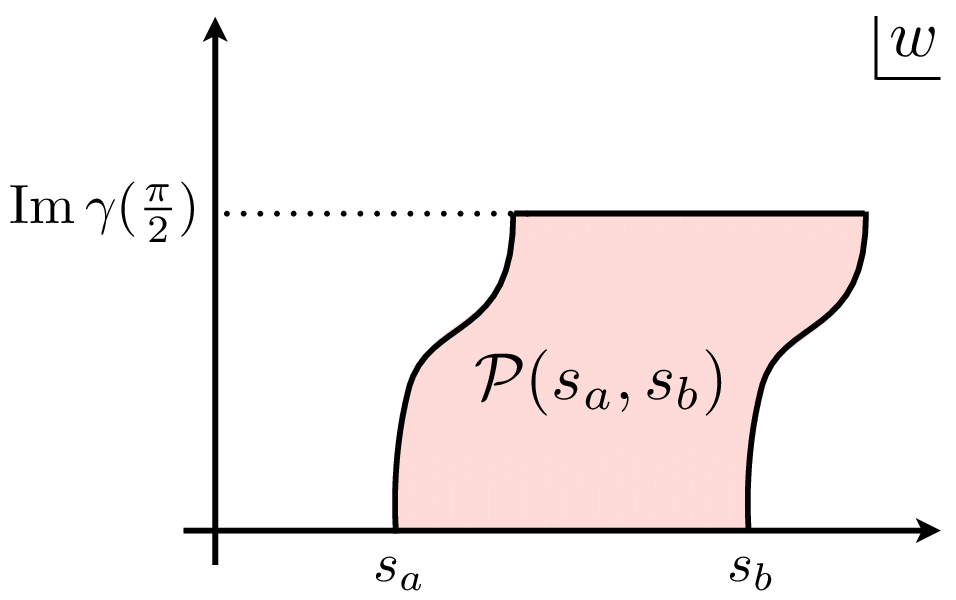}
  			\end{center}
  				\caption{A half propagator strip $\P(s_a, s_b)$.
				The edge on the real axis is the open string boundary,
				and that on Im $w={\rm Im}\,\gamma(\tfrac{\pi}{2})$ is the trajectory of the open string midpoint.}
  		 	\label{fig:one}
	\end{figure}

Then,
we move on to the concrete construction.
As we mentioned above,
the open string propagator in the linear $b$ gauge
is generated by
\begin{equation}
\mathcal{L}=\{Q,\,\lB\}
\quad
{\rm with}
\quad
\lB=\oint\frac{d\xi}{2\pi i}v(\xi)b(\xi)
=\oint\frac{d\xi}{2\pi i}\frac{\mathcal F_{\rm lin}(\xi)}{\mathcal F_{\rm lin}'(\xi)}b(\xi)\,,
\end{equation}
where we introduced a function $\mathcal F_{\rm lin}(\xi)$
given by
$\displaystyle v(\xi)=\frac{\mathcal F_{\rm lin}(\xi)}{\mathcal F^\prime_{\rm lin}(\xi)}$
and
$\mathcal F_{\rm lin}(1)=1$.
It is useful to define two coordinates 
$w$ and $z$ by
\begin{align}
e^w=2z=\mathcal F_{\rm lin}(\xi)\,.
\end{align}
In these frames,
the operator $\mathcal L$
generates a translation along the real axis
and a dilatation,
respectively, for the $w$-frame and the $z$-frame.
To define the half propagator strip,
we use
the 
 operator $\mathcal{L}_R(t)$ defined in the $w$-frame by
	\begin{align}
        	\mathcal{L}_R(t) \equiv \int_t^{\gamma(\frac{\pi}{2}) + t} \left[ \frac{dw}{2 \pi i}T(w) + \frac{d\bar{w}}{2 \pi i}\tilde{T}(\bar{w})  \right],\label{right_L}
        \end{align}
where $t$ is 
a real number and
the integral over $w$
is along the curve $\gamma(\theta)$ $(0\le \theta\le \pi)$:
\begin{equation}\label{wz-relation}
\gamma(\theta)=w(\xi=e^{i\theta})\,.
\end{equation}
We then define the half propagator strip in $w$ frame as follows:
	\begin{align}
        	\P(s_a, s_b) = \mathrm{P}\exp\left[ -\int^{s_b}_{s_a}dt\mathcal{L}_R(t) \right],
        \end{align}
where
$\mathrm{P}\exp$ denotes the path-ordered exponential
and the parameters $s_a$ and $s_b$ describe
positions of the strip on the real axis
(see Figure \ref{fig:one}). 

\medskip
Using the half propagator strip,
we construct the boundary state $|B\rangle$ for the original background as follows:
	\begin{align}
        	|B\rangle = e^{\frac{\pi^2}{s}(L_0 + \tilde{L}_0)}\oint_s \P(0,s)\,,
        \end{align}
where the operation $\oint_s$ denotes an identification of the left and right boundaries of the surface $\P(0,s)$,
which is realized as $w\sim w+s$ in the $w$-frame
and $ z \sim e^s z$ in the $z$-frame.
The operator~$\displaystyle e^{\frac{\pi^2}{s}(L_0 + \tilde{L}_0)}$
generates a propagation of the closed string
from the closed string boundary to the open string boundary.

\medskip
Next, we construct
the closed string state $|B_*(\Psi) \rangle$
by replacing $\mathcal{L}$ with $\mathcal{L}_*$.
The operator $\mathcal{L}_R(t)$ is then replaced by $\mathcal{L}_R(t) + \{ \B_R(t), \Psi \}$,\footnote{
It is because the action of the operator $\{Q_*,\lB\}$ on a string field $A$ is given by 
	\begin{align} 
	\no
        	 \{Q_*,\lB\} A = \left[\mathcal{L}_R A + (-)^A(\B_R A)\Psi - (-)^A\B_R(A \Psi)  \right] + \left[ \mathcal{L}_L A + \Psi (\B_L A) + \B_L (\Psi A)  \right],
        \end{align}
where $\mathcal{L}_{L/R}$ and $\B_{L/R}$ represent the left/right half of the line integrals $\mathcal{L}$ and $\lB$.
}
and the half propagator strip is also replaced as
	\begin{align}
        	\P(s_a,s_b) \rightarrow \P_*(s_a,s_b) \equiv \mathrm{P}\exp\left[ -\int^{s_b}_{s_a}dt \left( \mathcal{L}_R(t) + \{ \B_R(t),\Psi \} \right) \right].
        \end{align}
Using this deformed half propagator strip $\P_*(s_a,s_b)$,
we construct the closed string state $| B_*(\Psi) \rangle$ around the new background in analogy with $|B\rangle$:
 	\begin{align}
        	| B_*(\Psi) \rangle = e^{\frac{\pi^2}{s}(L_0 + \tilde{L}_0)}\oint_s \P_*(0,s).
        \end{align}
Expanding the path-ordered exponential in $\P_*(0,s)$, we can write the closed string state $| B_*(\Psi) \rangle$
as a power series in the classical solution
$\Psi$:
	\begin{align}
        	&|B_*(\Psi) \rangle\no\\
                =& \sum_{k=0}^{\infty} | B_*^{(k)}(\Psi)\rangle \no \\
                =& \sum_{k=0}^{\infty} (-)^ke^{\frac{\pi^2}{s}(L_0+\tilde{L}_0)}\oint_s \int_0^s ds_1 \cdots \int_{s_{i-1}}^s ds_i \cdots \int_{s_{k-1}}^s ds_k \P(0,s_1)\{ \B_R(s_1) ,\Psi  \}\P(s_1,s_2)\no\\
                &\qquad\qquad \times\cdots \P(s_{i-1},s_i)\{ \B_R(s_i) ,\Psi  \}\P(s_i,s_{i+1})\cdots \P(s_{k-1},s_k)\{ \B_R(s_k) ,\Psi  \}\P(s_k,s)\,,\label{general_boundary_state}
        \end{align}
where $|B_*^{(0)}(\Psi)\rangle = |B\rangle$. 
Products of half propagators and solutions
are defined by some gluing conditions,
which we explain in the next subsection
for half propagators in the Schnabl gauge. 
In the following,
we use this expression
to construct boundary states from classical solutions.

\subsubsection{Schnabl gauge calculation}
In general,
it is difficult to construct boundary states
from wedge-based classical solutions
because of the complicated gluing operation of
half propagator strips and classical solutions.
However,
it becomes tractable
when we use half propagator strips associated with Schnabl gauge propagators.\footnote{
The Schnabl gauge is singular
in the sense that the propagator
does not generate midpoint propagation.
Therefore,
it should be understood as
a singular limit of one-parameter family of regular linear $b$ gauges~\cite{Kiermaier:2007jg}
when the midpoint propagation
plays an important role.
In the original paper~\cite{Kiermaier:2008qu},
the Schnabl gauge calculation of boundary states is introduced
in this way,
and it is justified for classical solutions
constructed from wedge-based states of finite width
~\cite{Kiermaier:2008jy}.
Although
it is not clear whether
it is justified
when classical solutions contain
wedge-based states of infinite width,
we expect that
the use of the Schnabl gauge calculation
is justified when
our regularity conditions are satisfied.
} 

\begin{figure}[t]
\begin{center}
\includegraphics[width=170mm]{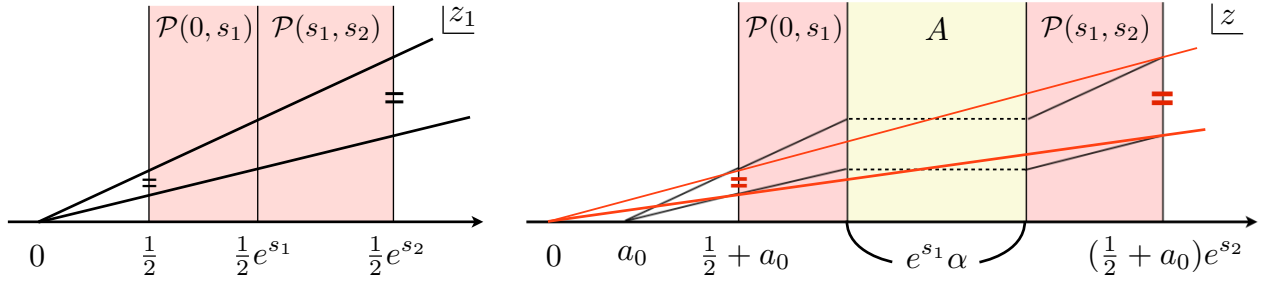}
 \end{center}
 \caption{
 The operation $\oint_{s_2}$ on the half propagator
$\P(0,s_1)\P(s_{1},s_{2})$
is realized by the identification $z_1\sim e^{s_2}z_1$ (the left figure).
After inserting a string field $A$,
the operation $\oint_{s_2}$ on $\P(0,s_1)\,A\,\P(s_{1},s_{2})$
is complicatedly realized in the $z_1$ frame.
However,
in the natural $z$ frame,
it is naturally realized
by $z\sim e^{s_2} z$ (the right figure).
}
 \label{natural_z}
\end{figure}

\medskip
In the Schnabl gauge calculation,
it is useful to perform in the $z$-frame,
which coincides with the sliver frame $z=\frac{2}{\pi}\arctan \xi$. In this frame,
the half propagator strip $\P(s_{i-1},s_i)$
can be described as a surface in the region
	\begin{align}
        	\frac{1}{2}e^{s_{i-1}} \leq {\rm Re}\, z_i \leq \frac{1}{2}e^{s_i}\,.
        \end{align}
Then,
let us consider an insertion of a wedge-based state $A$ of width~$\alpha$
between two half propagator strips
$\P(0,s_1)$ and $\P(s_1,s_2)$.
Associated with the fact that
the open string propagator generates
dilatation in the $z$-frame,
the gluing condition of $\P(0,s_1)$
and $A$ is given~by
\begin{align}
z_1=e^{s_1}z_{A}\,,
\end{align}
where $z_A$ is the coordinate of $A$ in the sliver frame.
In the same way,
the gluing condition of $A$ and $\P(s_{1},s_2)$
is given by
\begin{align}
e^{s_1}(z_A-\alpha)=z_{2}\,.
\end{align}
When we describe the surface $\P(0,s_1)\,A\,\P(s_{1},s_{2})$
in the $z_1$-frame,
it is located in the region
	\begin{align}
        	\frac{1}{2} \leq {\rm Re}\, z_1 \leq e^{s_1}\alpha + \frac{1}{2}e^{s_2}\,.
        \end{align}
To restore the fact that $\P(0,s_1)\P(s_{1},s_{2})=\P(0,s_{2})$
generates a dilatation $z\to e^{s_2}z$,
it is useful to introduce the following natural $z$-frame
(see Figure~\ref{natural_z}):
\begin{align}
z=z_1+a_0\quad{\rm with}\quad a_0=\frac{e^{s_1}}{e^{s_2}-1}\alpha\,.
\end{align}
In this frame,
the surface is located in the region
	\begin{align}
        	\frac{1}{2}+a_0 \leq {\rm Re}\, z \leq e^{s_2}\Big(\frac{1}{2}+a_0\Big)\,,
        \end{align}
and the operation $\oint_{s_2}$ on $\P(0,s_1)\,A\,\P(s_{1},s_{2})$
can be realized by the identification $z\sim e^{s_2}z$.

\medskip
Above discussion can be generalized straightforwardly
to the case with multi insertions of wedge-based states:
\begin{align}
\P(0,s_1)\,A_{1}\,\P(s_{1},s_{2})\,A_{2}\,\P(s_{2},s_{3})\,\ldots
\P(s_k-1,s_k)\,A_{k}\,\P(s_{k},s)\,,
\end{align}
where $A_i$ is a wedge-based state of width $\alpha_i$.
In the $z_1$-frame,
it is located in the region
	\begin{align}
        	\frac{1}{2} \leq {\rm Re}\, z_1 \leq \sum_{j=1}^{k} e^j \alpha_j + \frac{1}{2}e^s\,,
        \end{align}
and we define the natural $z$-frame by
\begin{align}
z=z_1+a_0\quad{\rm with}\quad a_0 = \frac{1}{e^s - 1}\sum_{j=1}^k e^{s_j}\alpha_j\,.
\end{align}
In the natural $z$-frame,
the surface is located in the region
	\begin{align}
        	\frac{1}{2}+a_0 \leq {\rm Re}\, z \leq e^{s_2}\Big(\frac{1}{2}+a_0\Big)\,,
        \end{align}
and the operation $\oint_{s}$ can be realized
by the identification $z\sim e^{s}z$.
We also note that
the conformal mapping between
the coordinate $z$ of the natural $z$-frame
and the coordinate $z_{A_i}$ of $A_i$ in the sliver frame
is given by
\begin{align}
z = \ell_i + e^{s_i}z_{A_i}\quad
{\rm with}
\quad
\ell_i = \sum_{j=1}^{i-1}\alpha_j e^{s_j} + a_0
\quad
{\rm and}
\quad
 \ell_1 = a_0\,.
\end{align}

\medskip
Then,
let us apply the above discussion to the boundary state~(\ref{general_boundary_state}).
When the classical solution $\Psi$
is constructed from wedge-based states,
$\{ \B_R(s_i), \Psi \}$ in~(\ref{general_boundary_state})
can be written as a sum of
states in the form of $\{ \B_R(s_i), A_{i} \}$,
where $A_{i}$ is a wedge-based state of width $\alpha_i$.
In the natural $z$ frame, this commutator is expressed as
	\begin{align}
        	-\{ \B_R(s_i), A_{\alpha_i} \} \rightarrow 
                \oint \frac{dz}{2 \pi i}(z - \ell_i)b(z)[\cdots] - e^{s_i}\alpha_i [\cdots]\B, \label{formula1}
        \end{align}
where $[\cdots]$ represents the operator insertions coming from $A_{i}$
and $\B$ is a line integral defined~by
\begin{align}
\B=\int_{i\infty}^{-i\infty}\frac{dz}{2\pi i}b(z)\,.
\end{align}
Then,
the calculation of boundary states is reduced to
the evaluation of
quantities in the form of $\B [\cdots]$.
Since the gluing operation $\oint_s$
leads to the identification $z\sim e^sz$ in the natural $z$-frame,
we obtain
\begin{align}
\B [\cdots]=e^s(-1)^{[\cdots]}[\cdots]\B\,,
\end{align}
where $[\cdots]$ represents all the insertions of operators on the surface.
Therefore,
we conclude that
        \begin{align}
                \B[\cdots] = \frac{e^s}{e^s - 1}\oint \frac{dz}{2 \pi i}b(z)[\cdots]\,, \label{formula2}
        \end{align}
where the contour encircles all the operator insertions $[\cdots]$ counterclockwise.
In the following section,
we use the formulae (\ref{formula1}) and (\ref{formula2})
to calculate boundary states in the $KBc$ subalgebra.

\subsection{Boundary states from string fields in $KBc$ subalgebra}
In this subsection we calculate
the closed string state $|B_*(\Psi) \rangle$
defined by~(\ref{general_boundary_state})
for general string fields $\Psi$
in the $KBc$ subalgebra:
\begin{equation}
\Psi=\sum_iF_i(K) cB G_i(K) c H_i(K)\,.
\label{general_KBc_index}
\end{equation}
As we mentioned in section~\ref{section_setup},
we assume that
$F_i(K)$, $G_i(K)$, and $H_i(K)$
are superpositions of wedge states,
and
they satisfy the regularity conditions introduced in section~\ref{subsection_regularity}.
We note that
the closed string state $|B_*(\Psi) \rangle$
is well-defined, although
it does not satisfy some properties of boundary states
unless $\Psi$ satisfies the equation of motion.

\medskip
We first rewrite
$\{ \B_R(s_i), \Psi \}$ in~(\ref{general_boundary_state})
using the formula~(\ref{formula1}).
Using the expression of $\Psi$
as a superposition of wedge-based states,
	\begin{align}
        	\Psi &= \sum_{i} \int_0^\infty d\alpha_i f_i(\alpha_i) \int_0^\infty d\beta_i g_i(\beta_i) \int_0^\infty d\gamma_i h_i(\gamma_i)
                         e^{\alpha_i K} cB e^{\beta_i K} c e^{\gamma_i K}\,,\label{general_field}
        \end{align}
the operator insertions in the natural $z$-frame
coming from each integrand
$e^{\alpha_i K} cB e^{\beta_i K} c e^{\gamma_i K}$
are given by
\begin{align}
e^{-s_i}c((\tfrac{1}{2}+\alpha_i)e^{s_i} + \ell_i))\,\mathcal{B}\,c((\tfrac{1}{2}+\alpha_i + \beta_i)e^{s_i} + \ell_i)\,.
\end{align}
Using the formula~(\ref{formula1}),
the operator insertions
coming from
$-\{ \B_R(s_j),e^{\alpha_i K} cB e^{\beta_i K} c e^{\gamma_i K}\}$
are
	\begin{align}
                        \left(\frac{1}{2} + \alpha_i \right)\B c((\tfrac{1}{2}+\alpha_i + \beta_i)e^{s_i} + \ell_i)
                           +\left( \frac{1}{2} - \gamma_i \right) c((\tfrac{1}{2}+\alpha_i)e^{s_j} + \ell_j)\B\,.				
                           \label{cBBc}			    
        \end{align}
We notice that
the first term and the second term
in~(\ref{cBBc})
are schematically in the form of $\B c$ and $c\B$, respectively.

\medskip
We then consider the operator insertions
in~$| B_*^{(k)}(\Psi)\rangle$
defined by~(\ref{general_boundary_state}).
Let us start from qualitative discussion.
As we mentioned above,
we have $\B c$-type and $c\B$-type
insertions from each $\{ \B_R(s_i), \Psi \}$.
Since $\B^2=0$,
cross terms between these two types vanish.
Then,
we have two types of operator insertions schematically in the form of
\begin{align}
\prod_{i=1}^k\Big(\B c(t_i)\Big)\quad
{\rm and}
\quad
\prod_{i=1}^k\Big( c(t_i)\B\Big)\,.
\end{align}
Furthermore,
using $\{\B,c(t_i)\}=1$ and the formula~(\ref{formula2}),
they can be written as
\begin{align}
\prod_{i=1}^k\Big(\B c(t_i)\Big)=\B c(t_k)=\frac{e^s}{e^s-1}\quad
{\rm and}
\quad
\prod_{i=1}^k\Big( c(t_i)\B\Big)= c(t_1)\B=-\frac{1}{e^s-1}\,,
\label{multi_insertions_Bc}
\end{align}
where we notice that
(\ref{multi_insertions_Bc}) does not depend on the insertion points
of $c$ ghosts.
Therefore,
we can ignore the argument of $c$ ghosts,
and $-\{ \B_R(s_i), \Psi \}$ can be written as follows:
	\begin{align}
	&-\{ \B_R(s_i), \Psi \}\no\\
	\rightarrow&\sum_{i} \int_0^\infty d\alpha_i f_i(\alpha_i) \int_0^\infty d\beta_i g_i(\beta_i) \int_0^\infty d\gamma_i h_i(\gamma_i)
                       \Big[\left(\frac{1}{2} + \alpha_i \right)\B c
                           +\left( \frac{1}{2} - \gamma_i \right) c\B\Big]\no\\
                           &=x\, \B c 
                        +y\, c\B\,,
        \end{align}
where $x$ and $y$ are 
$c$-numbers given by
\begin{align}
\label{def_alpha}
x=\sum_iG_i(0)\left( \frac{1}{2}F_i(0)H_i(0) + F'_i(0)H_i(0) \right)\,,\\
\label{def_beta}
y=\sum_iG_i(0)\left( \frac{1}{2}F_i(0)H_i(0) - F_i(0)H'_i(0) \right)\,.
\end{align}
Here
the regularity conditions on string fields introduced in section~\ref{subsection_regularity}
guarantee the finiteness of $x$ and $y$.
In the same way,
the operator insertions
in~$| B_*^{(k)}(\Psi)\rangle$
can be written as
	\begin{align}
        	   (-)^k\prod_{i=1}^k\{ \B_R(s_i),\Psi \}
                \rightarrow (x\, \B c)^k+(y\, c\B)^k
                =x^k\B c+y^kc\B=\frac{e^s}{e^s-1}x^k-\frac{1}{e^s-1}y^k\,.
        \end{align}
We then obtain
	\begin{equation}
            |B_*^{(k)}(\Psi)\rangle
                =\frac{s^k}{k!}\left( \frac{e^s}{e^s-1}x^k
                        -
                         \frac{1}{e^s-1}y^k\right)|B\rangle,
        \end{equation}
where the factor ${s^k}/{k!}$ comes from 
	\begin{align}
        	\int_0^sds_1\int_{s_1}^sds_2 \cdots \int_{s_{k-1}}^s ds_k = \frac{s^k}{k!}\,.
        \end{align}
Therefore, the all order state $|B_*(\Psi)\rangle$ is\footnote{
When the index $i$ in~(\ref{general_KBc_index}) runs over one value,
our result reproduces that in the previous paper~\cite{Takahashi:2011wk}.
}
\begin{equation}
\label{B_ast}
| B_\ast(\Psi)\rangle=\sum_{k=0}^{\infty}|B_*^{(k)}(\Psi)\rangle
	                      =\frac{e^{(x+1) s}-e^{y s}}{e^s-1}|B\rangle\,.
\end{equation}
We emphasize that
the closed string state $| B_\ast(\Psi)\rangle$ is proportional to
the boundary state $|B\rangle$ for the original BCFT,
and the information about the sting field $\Psi$
only appears in the $c$-numbers $x$ and $y$.

\subsection{The $s$-dependence of boundary states}
\label{subsection_s-dep}
As we mentioned earlier,
the closed string state $| B_\ast(\Psi)\rangle$
does not depend on the parameter $s$
when $\Psi$ satisfies the equation of motion.
In this subsection
we explore which class of classical solutions can be
in the $KBc$ subalgebra
by requiring the $s$-independence of $| B_\ast(\Psi)\rangle$
as a necessary condition
to satisfy the equation of motion.

\medskip
By differentiating~(\ref{B_ast})
with respect to $s$,
we obtain that
\begin{align}
\label{s-derivative}
\frac{\partial}{\partial s}|B_\ast(\Psi)\rangle=\frac{1}{(e^s-1)^2}\left[
x e^{(2+x)s}-(1+x)e^{(1+x)s}+(1-y)e^{(1+y)s}+y e^{y s}
\right]|B\rangle\,.
\end{align}
We first notice that the $s$-dependence does not vanish
when none of the following conditions are satisfied:
$y=x$, $y=x+1$, and $y=x+2$.
Then,
let us consider the following three cases:
\begin{enumerate}
\item $y=x$

The $s$-derivative~(\ref{s-derivative}) is given by
\begin{equation}
\frac{\partial}{\partial s}|B_\ast(\Psi)\rangle=\frac{1}{(e^s-1)^2}\left[
x e^{(2+x)s}-2x e^{(1+x)s}+x e^{x s}
\right]|B\rangle\,,
\end{equation}
and it vanishes if and only if $x=0$.
For $x=y=0$,
the closed string state~(\ref{B_ast}) takes the form
\begin{equation}
|B_\ast(\Psi)\rangle=|B\rangle\,,
\end{equation}
and
the boundary state
for the perturbative vacuum can
be reproduced by this class
of string fields.

\medskip
\item $y=x+1$

The $s$-derivative~(\ref{s-derivative}) is identically zero:
\begin{equation}
\frac{\partial}{\partial s}|B_\ast(\Psi)\rangle=\frac{1}{(e^s-1)^2}\left[
x e^{(2+x)s}-(1+x)e^{(1+x)s}-x e^{(2+x)s}+(1+x) e^{(1+x) s}
\right]|B\rangle=0\,.
\end{equation}
For $y=x+1={\rm arbitrary}$,
the closed string state~(\ref{B_ast}) takes the form
\begin{equation}
|B_\ast(\Psi)\rangle=0\,,
\end{equation}
and
the boundary state for the tachyon vacuum
can be reproduced by
this class of string fields.

\medskip
\item $y=x+2$

The $s$-derivative~(\ref{s-derivative}) is given by
\begin{equation}
\frac{\partial}{\partial s}|B_\ast(\Psi)\rangle=\frac{1}{(e^s-1)^2}\left[
2(1+x) e^{(2+x)s}-(1+x)e^{(1+x)s}-(1+x)e^{(3+x)s}\right]|B\rangle\,,
\end{equation}
and it vanishes if and only if $x=-1$.
For $x=-1$ and $y=1$,
the closed string state~(\ref{B_ast}) takes the form
\begin{equation}
|B_\ast(\Psi)\rangle=-|B\rangle\,.
\end{equation}
This type of string fields reproduce
boundary states for the original BCFT with $-1$ factor,
which coincide with
that for the ghost D-brane introduced in~\cite{Okuda:2006fb}.
We call solutions reproducing such a boundary state ghost brane solutions.
In section~\ref{ghost_sec},
we give an example of this type of classical solutions.
\end{enumerate}

\bigskip
Summarizing above discussions,
the $s$-dependence of the closed string state $|B_\ast(\Psi)\rangle$ vanishes
only in the following three cases:
$x=y=0$,
$x=y-1={\rm arbitrary}$,
and $x=y-2=-1$.
They reproduce the boundary states
for the perturbative vacuum,
the tachyon vacuum,
and the ghost D-brane,
respectively.

\paragraph{-- Comments on singular string fields and their regularization}
Our results suggest
that we are not able to
construct multiple D-brane solutions
in the $KBc$ subalgebra
without relaxing our regularity conditions.
Indeed,
the proposed multiple D-brane solutions
in~\cite{Murata:2011ex,Murata:2011ep}
contain singular expression such as $1/K$,
and
it is not known how to regularize them consistently.
In the last of this section
we comment on regularization
of singular string fields.

\medskip
Suppose that
a string field $\Psi$ of ghost number $1$
contains a singular expression
such as $1/K$,
and we regularize it
using
a one-parameter family $\Psi_\epsilon$
of regular string fields satisfying
our regularity conditions:
\begin{equation}
\Psi=\lim_{\epsilon\to0}\Psi_\epsilon
\quad
{\rm with}
\quad
\Psi_\epsilon=\sum_iF^\epsilon_i(K) cB G^\epsilon_i(K) c H^\epsilon_i(K)\,.
\end{equation}
For $\Psi_\epsilon$ ($\epsilon\neq0$),
we can evaluate
the closed string state $|B_*(\Psi)\rangle$
using the general formula~(\ref{B_ast}):
\begin{align}
|B_*(\Psi_\epsilon)\rangle
=\left[\frac{e^s}{e^s-1}\exp (x_\epsilon s)-\frac{1}{e^s-1}\exp\left(y_\epsilon s\right)\right]|B\rangle\,,
\label{boundary_epsilon}
\end{align}
where $x_\epsilon$ and $y_\epsilon$
are the parameters (\ref{def_alpha}) and (\ref{def_beta})
for $\Psi_\epsilon$.
We notice that
the $s$-independence of boundary states
prohibits multiple D-brane solutions
as long as we formally take the limit $\epsilon\to0$
in the expression~(\ref{boundary_epsilon}).
We should mention,
however,
that we used the Schnabl gauge calculation
to derive our formula,
and it is not clear whether
the Schnabl gauge calculation is justified in the limit $\epsilon\to0$.
In section~\ref{discussion_sec}
we discuss what
we should do
to clarify this point.

\section{Constraints on pure-gauge ansatz}
\label{pure_sec}
\setcounter{equation}{0}
In this section
we apply the results in the previous section
to string fields formally in the pure-gauge form.
We show that this class of solutions
include only 
those reproducing the following boundary states:
$|B_*\rangle = |B\rangle$ and $0$.

\subsection{Pure-gauge ansatz in $KBc$ subalgebra}
\label{pure_geuge_ansatz}
Any string field $U$ of ghost number $0$
in the $KBc$ subalgebra
can be written~as
\begin{equation}
\label{U_in_KBc}
U=H(K)-\sum_{i}I_i(K)cB\,J_i(K)\,,
\end{equation}
where $H(K)$, $I_i(K)$, and $J_i(K)$
are arbitrary functions of the state $K$.
We formally define the inverse of $U$ by
\begin{equation}
U^{-1}=\Big[1-H^{-1}(K)\sum_{j}I_j(K)Bc\,J_j(K)\Big]\frac{1}{H(K)-\sum_{i}I_i(K)J_i(K)}\,,
\end{equation}
where we assumed that $H(K)\neq0$ and $H(K)-\sum_{i}I_i(K)J_i(K)\neq0$.
Then, 
the most general form of the pure-gauge ansatz
in the $KBc$ subalgebra is given by
\begin{eqnarray}
\nonumber
 \Psi&=&U^{-1}QU\\
&=&\sum_{i}H^{-1}I_icKBcJ_i+\sum_{i}H^{-1}I_icJ_i\frac{KB}{H-\sum_{j}I_jJ_j}\sum_{k}I_kcJ_k\,.
\label{general_solutions}
\end{eqnarray}
When $H(K)\neq0$,
we can set $H(K)=1$ without loss of generality,
and therefore,
we consider the following class of formal solutions
in the rest of this section:
\begin{equation}
\label{pure_gauge_ansatz}
\Psi=\sum_{i}I_icKBcJ_i+\sum_{i,j}I_icBG_{ij}cJ_j
\quad
{\rm with}
\quad
G_{ij}=\frac{KJ_iI_j}{1-\sum_kI_kJ_k}\,.
\end{equation}
Note that when 
the indices $i$ and $j$ in (\ref{pure_gauge_ansatz}) run over only one value,
we obtain Okawa's formal solutions~(\ref{Okawa's_solution}).
In the following,
we assume that
$I_i(K)$ and $J_i(K)$ satisfy the regularity conditions I and II
introduced in section~\ref{subsection_regularity},
and $G_{ij}(K)$ satisfies the regularity condition I.
We do not assume that $U$ and $U^{-1}$ are regular:
the formal solution $\Psi$ can be regular even if $U$ or $U^{-1}$ is singular
as is the case of Schnabl's original solution~\cite{Schnabl:2005gv,Okawa:2006vm}
and the simple solution~\cite{Erler:2009uj} for the tachyon vacuum.

\subsection{Constraints using boundary states}\label{4.2}
We then clarify which class of solutions are
in this class of formal solutions~(\ref{general_solutions})
using the property of boundary states.
Applying the general formulae,
we obtain that
\begin{eqnarray}
\label{alpha_for_pure}
x&=&\sum_{ij}\left[G_{ij}(0)\left(\frac{1}{2}I_i(0)J_j(0)+I_i^\prime(0)J_j(0)\right)\right]\,,\\
\label{beta_for_pure}
y&=&\sum_{ij}\left[G_{ij}(0)\left(\frac{1}{2}I_i(0)J_j(0)-I_i(0)J^\prime_j(0)\right)\right]\,,
\end{eqnarray}
where the parameters $x$ and $y$
are well-defined
because of the regularity conditions.
Then,
let us consider 
the following three cases.
\begin{enumerate}
 \item $\sum_j I_j(0) J_j(0)\neq1$
  
\smallskip
The regularity conditions imply
\begin{equation}
G_{ij}(0)=\frac{0 \cdot J_i(0) I_j(0)}{1-\sum_k I_k(0) J_k(0)}=0\,,
\end{equation}
which leads to $x=y=0$.
Therefore,
the boundary state for
the perturbative vacuum
can be reproduced by
this class of string fields:
$|B_*(\Psi)\rangle=|B\rangle$.
\smallskip
  \item $\sum_j I_j(0) J_j(0)=1\quad {\rm and}\quad
\sum_j\left(I_j^\prime(0) J_j(0) + I_j(0) J_j^\prime(0)\right)\neq 0$

\smallskip
The regularity conditions imply
\begin{align}
\sum_k I_k(z)J_k(z) = 1 + \sum_k\left(I_k^\prime(0) J_k(0) + I_k(0) J_k^\prime(0)\right) z + \mathcal{O}(z^2)\,,
\end{align}
which leads to
\begin{eqnarray}
G_{ij}(0)&=&-\frac{J_i(0)I_j(0)}{\sum_k\left[I^\prime_k(0)J_k(0)+I_k(0)J_k^\prime(0)\right]}\,.
\end{eqnarray}
We then obtain
\begin{eqnarray}
x&=&-\frac{\frac{1}{2}+\sum_iI^\prime_i(0)J_i(0)}
               {\sum_k\left[I^\prime_k(0)J_k(0)+I_k(0)J_k^\prime(0)\right]}\,,\\
y&=&-\frac{\frac{1}{2}-\sum_iI_i(0)J^\prime_i(0)}
              {\sum_k\left[I^\prime_k(0)J_k(0)+I_k(0)J_k^\prime(0)\right]}=x+1\,.
\end{eqnarray}
Therefore,
the boundary state for the tachyon vacuum
can be reproduced by
this class of string fields:
$|B_*(\Psi)\rangle=0$.
\smallskip
  \item $\sum_j I_j(0) J_j(0)=1\quad {\rm and}\quad
\sum_j\left(I_j^\prime(0) J_j(0) + I_j(0) J_j^\prime(0)\right)= 0$

\smallskip
The regularity condition on $G_{ij}(K)$ implies
$I_j(0)J_i(0)=0$ for any $i$ and $j$.
However,
it is incompatible with the assumption $\sum_jI_j(0)J_j(0)=1$.
Therefore,
no string fields in this class
satisfy our regularity conditions.
\end{enumerate}

\medskip
Summarizing above discussions,
the boundary states 
for the perturbative and the tachyon vacuum
can be reproduced by
the formal solutions
based on the pure-gauge ansatz in the $KBc$ subalgebra~(\ref{general_solutions}).
In particular,
the ghost brane solutions
are not in this class of formal solutions.

\subsection{Examples}
In this subsection
we consider two examples
for the pure-gauge ansatz.
We first consider the solutions for tachyon condensation~\cite{Schnabl:2005gv,Erler:2009uj},
whose boundary states are already evaluated in~\cite{Kiermaier:2008qu,Takahashi:2011wk}.
We make an observation that
only the so-called phantom term contributes to the boundary states
for solutions in the $KBc$ subalgebra.
We then consider the proposed multiple D-brane solutions~\cite{Murata:2011ex,Murata:2011ep}.
After formally applying our results to the proposed solutions,
we discuss two types of regularization~\cite{Murata:2011ep,Hata:2011ke} and \cite{Masuda} in our framework.

\subsubsection{Solutions for tachyon condensation
and the phantom term}
\label{subsubsub_phantom}
\paragraph{-- Schnabl's original solution}
We start from Schnabl's original solution for tachyon condensation~\cite{Schnabl:2005gv}:
\begin{align}
\Psi_{\rm Schnabl}&=\lim_{N\to\infty}\left[\sum_{n=0}^Ne^{K/2}cKBe^{nK}c\,e^{K/2}-e^{K/2}cBe^{NK}c\,e^{K/2}\right]\,,
\end{align}
where the second term is the so-called phantom term.
Applying our general formulae,
we notice that the first term does not contribute to
the parameters $x$ and $y$
because of the factor~$K$ in $cKBe^{nK}c$.
This is consistent with Okawa's observation~\cite{Okawa:2006vm}
that the first term is a pure-gauge solution by itself.
The phantom term makes a non-trivial contribution:
$(x,y)=(-1,0)$, and therefore,
$|B_\ast(\Psi_{\rm Schnabl})\rangle=0.$

\paragraph{-- Simple solution for tachyon condensation}
We then consider the simple solution for tachyon condensation~\cite{Erler:2009uj}:
\begin{align}
\Psi_{\rm ES}&=\frac{1}{\sqrt{1-K}}(cKBc-c)\frac{1}{\sqrt{1-K}}\,.
\end{align}
As discussed in~\cite{Erler:2009uj},
we can rewrite it as
\begin{align}
\Psi=\lim_{\lambda\to1^-}\Psi_\lambda
-\frac{1}{\sqrt{1-K}}cB\widetilde{\Omega}^\infty c\frac{1}{\sqrt{1-K}}\,,
\end{align}
where the second term is the phantom term.
The state $ \widetilde{\Omega}^\infty$ is defined by
$\widetilde{\Omega}^\infty=\lim_{\epsilon\to0}\frac{\epsilon}{\epsilon-K}$,
and $\Psi_\lambda$ is 
the following one parameter family
of pure-gauge solutions:
\begin{align}
\nonumber
\Psi_\lambda&=\Big(1-\lambda\frac{1}{\sqrt{1-K}}cB\frac{1}{\sqrt{1-K}}\Big)^{-1}Q\Big(1-\lambda \frac{1}{\sqrt{1-K}}cB\frac{1}{\sqrt{1-K}}\Big)\\
&=\lambda \frac{1}{\sqrt{1-K}}(cKBc-c)\frac{1}{\sqrt{1-K}}
+\lambda \frac{1}{\sqrt{1-K}}c(1-K)\frac{1-\lambda}{1-\lambda-K}Bc\frac{1}{\sqrt{1-K}}\,,
\end{align}
which is well-defined for $0\leq\lambda<1$.
Applying our general formulae,
we again notice that
the pure-gauge part $\displaystyle{\lim_{\lambda\to1^-}\Psi_\lambda}$
does not contribute
to the parameters $x$ and $y$,
and
only the phantom term
contributes:
$(x,y)=(-1,0)$, and therefore,
$|B_\ast(\Psi_{\rm ES})\rangle=0$.

\paragraph{-- More general discussion}
Finally,
let us consider more general situation.
In~\cite{Erler:2012qr},
Erler and Maccaferri gave an interpretation
to the phantom term.
For any non-trivial solution~$\Psi$,
we can define
a non-trivial singular left gauge transformation~$U$
connecting to the perturbative vacuum~\cite{Erler:2012qn}:
\begin{equation}
U\Psi=QU\,,
\end{equation}
where $U$ has a non-trivial kernel\footnote{
We regard a multiplication by the string field $U_{\widetilde{\Psi}\Psi}$
as a morphism from a set of string fields to themselves~\cite{Erler:2012qn}.} 
and its inverse is ill defined.
Suppose that
we can take a small number $\epsilon$
so that $\epsilon+U$ does not have any non-trivial kernel.\footnote{
For example, we take $\epsilon$ to be a negative real number
when $U=K$.}
Then, the inverse of $\epsilon+U$ is well defined,
and
we can rewrite $\Psi$
using $(\epsilon+U)^{-1}$ as follows~\cite{Erler:2012qn}:
\begin{align}
\Psi=(\epsilon+U)^{-1}Q(\epsilon+U)+\frac{\epsilon}{\epsilon+U}\Psi\,.
\end{align}
Taking the limit $\epsilon\to0$,
we obtain
\begin{align}
\label{phantom_general}
\Psi=\lim_{\epsilon\to0}(\epsilon+U)^{-1}Q(\epsilon+U)+X^\infty\Psi\,,
\end{align}
where $X^\infty=\lim_{\epsilon\to0}\displaystyle{\frac{\epsilon}{\epsilon+U}}$
is the projector onto the kernel of $U$
called the boundary condition changing projector.\footnote{
See appendix~\ref{app_pro} for more discussion on the projector
and associated property in the $KBc$ subalgebra.}
Erler and Maccaferri gave an interpretation that
the second term in (\ref{phantom_general})
can be understood as the phantom term for known solutions
and
they showed its usefulness in the calculation of the energy
and the gauge-invariant observables.

\medskip
As we discussed,
the phantom term of the solution for tachyon condensation~\cite{Schnabl:2005gv,Erler:2009uj}
determines the property of the boundary states.
More generally in the $KBc$ subalgebra,
it is obvious from the expression (\ref{def_alpha}) and (\ref{def_beta})
that
the phantom term defined by $X^\infty\Psi$
determines the 
form
of the boundary states:
The parameters $x$ and $y$
are linear combinations of the contribution from each term of the solution.
The first term in~(\ref{phantom_general})
is a pure-gauge solution by itself
and it does not contribute to $x$ and $y$.
Then, only the phantom term contributes to the boundary states.
Let us discuss the case of Okawa's formal solutions for example:
\begin{align}
\label{Yuji_tachyon}
\Psi&=F(K)c\frac{KB}{1-F(K)^2}cF(K)\,,
\end{align}
where we assume that $F(0)^2=1$
so that $\Psi$ is not a pure-gauge solution.
Since~(\ref{Yuji_tachyon}) can be written as 
\begin{align}
U\Psi=QU\quad{\rm with}\quad
U=1-F(K)cBF(K)\,,
\end{align}
we obtain
\begin{align}
\Psi=F(K)c\frac{KB}{1+\epsilon-F(K)^2}cF(K)
+F(K)cB\frac{K}{1-F(K)^2}\frac{\epsilon}{\epsilon+1-F(K)^2}cF(K)\,,
\end{align}
where the first term is a pure-gauge solution by itself
and the second term in the limit $\epsilon\to0$ is the phantom term.
Applying our general formulae,
we notice that the first term does not contribute
to $x$ and $y$
and only the second term contributes.
It is also obvious that
the contribution from the second term
coincides with that from the solution (\ref{Yuji_tachyon}) itself
because $\displaystyle{\lim_{z\to0}}\frac{z}{1-F(z)^2}\frac{\epsilon}{\epsilon+1-F(z)^2}=\displaystyle{\lim_{z\to0}}\frac{z}{1-F(z)^2}$.

\medskip
Summarizing above discussion,
the boundary states for solutions in the $KBc$ subalgebra
are determined only from the phantom term $X^\infty\Psi$.
Since our discussion
highly depends on the expression of the formulae~(\ref{def_alpha}),
(\ref{def_beta}),
and (\ref{B_ast}) in the $KBc$ subalgebra,
it is not clear what is the situation for more general solutions
such as
marginal solutions~\cite{Kiermaier:2007ba,Kiermaier:2007vu,Kiermaier:2010cf}.
It would be interesting to discuss
the role of the phantom term
in the calculation of the boundary states
for them.

\subsubsection{Proposed multiple D-brane solutions and their regularization}
We then consider the multiple D-brane solutions proposed in~\cite{Murata:2011ex,Murata:2011ep},
which are in the class of Okawa's formal solutions:\footnote{
It is straightforward to extend the following discussion
to more general formal solutions~(\ref{general_solutions}).
However,
we concentrate on the Okawa's formal solutions for simplicity.}
\begin{align}
\label{Okawa_multi}
\Psi&=F(K)c\frac{KB}{1-F(K)^2}cF(K)\,.
\end{align}
Although the formal solution
(\ref{Okawa_multi}) does not satisfy the
regularity conditions in general, 
formally applying the general formulae,
we obtain
\begin{align}
\label{formal_alpha}
x&=\lim_{z\to0}\left[\frac{1}{2}\frac{zF(z)^2}{1-F(z)^2}-\frac{1}{2}\frac{z\big(1-F(z)^2\big)^\prime}{1-F(z)^2}\right]\,,\\
\label{formal_beta}
y&=\lim_{z\to0}\left[\frac{1}{2}\frac{zF(z)^2}{1-F(z)^2}+\frac{1}{2}\frac{z\big(1-F(z)^2\big)^\prime}{1-F(z)^2}\right]\,.
\end{align}
First,
let us consider the state $|B_\ast(\Psi)\rangle$
in the limit $s\to0$:
\begin{equation}
|B_\ast(\Psi)\rangle=(1+x-y)|B\rangle+\mathcal{O}(s)=\left(1-\lim_{z\to0}\frac{z\big(1-F(z)^2\big)^\prime}{1-F(z)^2}\right)|B\rangle+\mathcal{O}(s)\,.
\end{equation}
As discussed in~\cite{Takahashi:2011wk},
the state in the limit $s\to0$ reproduces the boundary state for $n$ D-branes
when the function $F(z)$ satisfies the following property:
\begin{equation}
\label{n_condition}
\lim_{z\to0}z\frac{\left(1-F(z)^2\right)^\prime}{1-F(z)^2}=1-n
\quad
{\rm or}
\quad
1-F(z)^2=a\, z^{1-n}+\ldots\,,
\end{equation}
where $a$ is a non-zero constant
and the dots stand for higher order terms in $z$.
This result in the limit $s\to0$
is essentially equivalent to
the calculation of
the gauge-invariant observables in~\cite{Murata:2011ex,Murata:2011ep}.
However,
the situation is different for finite $s$.
When $F(z)$ satisfies
(\ref{n_condition}),
the parameters $x$ and $y$ are given~by
\begin{equation}
\label{n_general}
(x,y)=\left\{\begin{array}{cl}
\displaystyle{\Big(\frac{n-1}{2},-\frac{n-1}{2}\Big)}&{\rm for}\quad n>0\,, \\[3mm]
\displaystyle{\Big(a-\frac{1}{2},a+\frac{1}{2}\Big)}& {\rm for}\quad n=0\,,\\[3mm]
({\rm singular},{\rm singular})&{\rm for}\quad n<0\,.\end{array}\right.
\end{equation}
Since the closed string state $|B_\ast(\Psi)\rangle$
has a non-trivial $s$-dependence
unless $x=y=0$ or $x=y-1$,
that
for (\ref{n_general})
does not reproduce an appropriate boundary state
when $n\neq0,1$.

\medskip
In the following,
we try to improve this formal observation
using two types of concrete regularization.\footnote{
As we mentioned in section~\ref{subsection_s-dep},
it is not clear whether
the Schnabl gauge calculation is justified in the singular limit.
However,
in this subsection,
we formally apply our results based on
the Schnabl gauge calculation
in order to discuss the relation to
the regularizations in~\cite{Murata:2011ep,Hata:2011ke,Masuda}
of proposed multiple D-brane solutions.}
We mainly consider the regularization for the following
double brane solution
for simplicity:
\begin{equation}
\label{singular_double}
\Psi_{\rm double}=-\frac{1}{\sqrt{-K}}c\frac{K^2}{K-1}Bc\frac{1}{\sqrt{-K}}\,,
\end{equation}
where we chose $F(K)=1/\sqrt{-K}$.

\paragraph{-- $K_\epsilon$ regularization}
We first consider the so-called $K_\epsilon$ regularization
discussed in~\cite{Murata:2011ep,Hata:2011ke}:
\begin{equation}
\label{epsilon_reg}
\Psi_{\rm HKMS}=-\lim_{\epsilon\to 0}\frac{1}{\sqrt{\epsilon-K}}c\frac{(K-\epsilon)^2}{K-\epsilon-1}Bc\frac{1}{\sqrt{\epsilon-K}}\,,
\end{equation}
where all $K$'s in~(\ref{singular_double}) are replaced by $K-\epsilon$.
In~\cite{Murata:2011ep,Hata:2011ke},
it was shown that
(\ref{epsilon_reg}) reproduces the energy
and the gauge-invariant observables for double branes.
It also satisfies the equation of motion
when contracted to the solution itself.
However,
the equation of motion is not satisfied
when contracted to some states in the Fock space\cite{Murata:2011ep}.

\medskip
Then,
let us calculate the boundary states
using $K_\epsilon$~regularization.
We consider the $K_\epsilon$~regularization
in the following general setting:
\begin{align}
\Psi=\lim_{\epsilon\to0}\Psi_\epsilon
\quad{\rm with}\quad
\Psi_\epsilon=F(K-\epsilon)c\frac{(K-\epsilon)B}{1-F(K-\epsilon)^2}cF(K-\epsilon)\,.
\end{align}
Applying our general formulae,
we obtain
\begin{align}
x_\epsilon&=\frac{1}{2}\frac{(-\epsilon)\, F(-\epsilon)^2}{1-F(-\epsilon)^2}-\frac{1}{2}\frac{(-\epsilon)\,\big(1-F(-\epsilon)^2\big)^\prime}{1-F(-\epsilon)^2}\,,\\
y_\epsilon&=\frac{1}{2}\frac{(-\epsilon)\,F(-\epsilon)^2}{1-F(-\epsilon)^2}+\frac{1}{2}\frac{(-\epsilon)\,\big(1-F(-\epsilon)^2\big)^\prime}{1-F(-\epsilon)^2}\,.
\end{align}
Here,
these expressions
coincide with those in the formal discussion
(\ref{formal_alpha}) and (\ref{formal_beta}).
Therefore,
the calculation of boundary states in the $K_\epsilon$ regularization
results in the formal discussion (\ref{n_general}).
For the double brane solution~(\ref{epsilon_reg}),
the state $|B_\ast(\Psi)\rangle$ is given by
\begin{equation}
|B_\ast(\Psi)\rangle=\lim_{\epsilon\to0}|B_\ast(\Psi_\epsilon)\rangle
=\big(e^{\frac{s}{2}}+e^{-\frac{s}{2}}\big)|B\rangle
=2|B\rangle+\mathcal{O}(s)\,.
\end{equation}
This result is consistent with the results in~\cite{Murata:2011ep}:
The non-trivial $s$-dependence implies
the violation of the equation of motion.
The calculation of the state $|B_\ast(\Psi)\rangle$ in the $s\to0$ limit,
which is essentially equivalent to that of the gauge-invariant observables,
reproduces the boundary states for double branes.

\medskip
It is also notable that
$\Psi_\epsilon$ can be written as follows:
\begin{align}
\Psi_\epsilon=F(K-\epsilon)c\frac{KB}{1-F(K-\epsilon)^2}cF(K-\epsilon)-F(K-\epsilon)c\frac{\epsilon B}{1-F(K-\epsilon)^2}cF(K-\epsilon)\,,
\end{align}
where the first term is a pure-gauge solution by itself
and the second term looks like a phantom term.
Although the second term leads to the violation of the equation of motion
for double brane solutions,
it reproduces the correct phantom term for solutions for tachyon condensation.
For example,
the second term for the simple solution for tachyon condensation
is given by
\begin{align}
({\rm second\,\,term})=-\frac{1}{\sqrt{1+\epsilon-K}}c(1+K-\epsilon)\frac{\epsilon}{\epsilon-K}Bc\frac{1}{\sqrt{1+\epsilon-K}}\,.
\end{align}
In the limit $\epsilon\to0$,
it reproduces the correct phantom term:\footnote{
We ignore the $1+\epsilon-K$ factor
since this would give a subleading contribution to the phantom piece
as discussed in~\cite{Erler:2009uj}.}
\begin{align}
\nonumber
({\rm second\,\,term})&\to-\lim_{\epsilon\to0}\left[\frac{1}{\sqrt{1+\epsilon-K}}c(1+\epsilon-K)\frac{\epsilon}{\epsilon-K}Bc\frac{1}{\sqrt{1+\epsilon-K}}\right]\\
&=-\frac{1}{\sqrt{1-K}}c\,\widetilde{\Omega}^\infty Bc\frac{1}{\sqrt{1-K}}\,.
\end{align}
Then,
the same discussion as
that in section~\ref{subsubsub_phantom} holds.

\paragraph{-- Regularization in~\cite{Masuda}}
We then consider another regularization discussed in~\cite{Masuda}
by one of the authors 
for the double brane solution~(\ref{singular_double}):
\begin{equation}
\label{Masuda_double}
\Psi_{\rm TM}=\Psi_{R_0}-\varphi_p\,,
\end{equation}
 where
  \begin{align}\label{cut4}
 \Psi_{R_0}&=-\lim_{\Lambda\to\infty }\int_0^\infty R_0(\Lambda;\,x) e^{Kx}dx\,c\frac{K^2}{K-1}Bc
 \quad
 {\rm with}
 \quad
 R_0(\Lambda;\, x)=1-\frac{\ln (x+1)}{\ln (\Lambda+1)}\,,
 \end{align}
 and $\varphi_p $ is a correction term whose expression is 
presented in \cite{Masuda}. 
In~\cite{Masuda},
it was shown that
the equation of motion for~(\ref{Masuda_double})
is satisfied
when contracted to
the state in the Fock space
and the solution itself,
and the solution~(\ref{Masuda_double}) reproduces the expected energy
for double branes.
However,
it was also shown that
the gauge-invariant observables~\cite{Ellwood:2008jh}
for~(\ref{Masuda_double})
are not those for the double brane background
but those for the perturbative vacuum.

\medskip
Then,
let us calculate the boundary states using this regularization.
Applying our general formulae,
we obtain
\begin{equation}
x=y=0\,,
\end{equation}
and therefore,
\begin{equation}
|B_\ast\rangle=|B\rangle\,,
\end{equation}
which is the boundary state for the perturbative vacuum.
Although the solution~(\ref{Masuda_double}) does not
reproduce the boundary state for double branes,
our result is not inconsistent with the results in~\cite{Masuda}:
The boundary state is $s$-independent
while the solution satisfies the equation of motion.
Both of the boundary state and the gauge-invariant observables
reproduce those for the perturbative vacuum.
However,
it is mysterious that the solution~(\ref{Masuda_double})
reproduces the boundary state and the gauge-invariant observables
for the perturbative vacuum while it reproduces
the energy for double branes.

\section{Another class of formal solutions}
\label{ghost_sec}
\setcounter{equation}{0}
In the previous section
we showed that the formal solutions based on the pure-gauge ansatz~(\ref{pure_gauge_ansatz})
do not
reproduce the boundary state for the ghost brane solution.
In this section
we discuss another class of formal solutions.

\medskip
Suppose that $\Psi$
is
a linear combination of two 
Okawa's formal solutions $\Psi_1$ and $\Psi_2$:
\begin{equation}
\Psi=\Psi_1+\Psi_2
\end{equation}
with
\begin{equation}
\Psi_1=F(K)c\frac{KB}{1-F(K)^2}cF(K)\,,
\quad
\Psi_2=H(K)c\frac{KB}{1-H(K)^2}cH(K)\,.
\end{equation}
Formally using the equations of motion for $\Psi_1$ and $\Psi_2$,
the equation of motion for $\Psi$
is reduced~to
\begin{equation}
\label{eom_ghost_1}
Q\Psi+\Psi^2=Q\Psi_1+\Psi_1^2+Q\Psi_2+\Psi_2^2+\Psi_1\Psi_2+\Psi_2\Psi_1
=\Psi_1\Psi_2+\Psi_2\Psi_1\,,
\end{equation}
and the cross terms in~(\ref{eom_ghost_1}) vanish
when $F(K)H(K)=1$ because $c^2=0$.\footnote{
Here we can alternatively take $F(K)H(K)=a$, where $a$ is an arbitrary nonzero constant. However, the resulting solution $\Psi_a$ for $a\neq1$, 
\begin{equation}
\label{pure_another}
\Psi_a=F(K)c\frac{KB}{1-F(K)^2}cF(K)+(aF(K)^{-1})c\frac{KB}{1-(aF(K)^{-1})^2}c(aF(K)^{-1})\,,\end{equation}
{\it can} be written in
the form~(\ref{pure_gauge_ansatz}).
}
We then obtain
the following class of formal solutions:
\begin{equation}
\Psi=F(K)c\frac{KB}{1-F(K)^2}cF(K)+H(K)c\frac{KB}{1-H(K)^2}cH(K)
\label{ghost_brane}
\end{equation}
with
\begin{equation}
F(K)H(K)=1\,,
\end{equation}
where
we assume that $F(K)$ and $H(K)$
satisfy the regularity conditions I and II
introduced in section~\ref{subsection_regularity},
and $K/(1-F^2)$
and $K/(1-H^2)$
satisfy the regularity condition I.
Here we note that
this class of formal solutions cannot be
written in the pure-gauge form~(\ref{pure_gauge_ansatz})
as discussed in appendix~\ref{appendix_B}.

\medskip
We also note that
the formal solution \eqref{ghost_brane}
seems to contain some identity-based terms:
For simplicity, let $F(K)^2$ be a meromorphic function of $K$.
Then, either $F^2$ or $1/F^2$ inevitably becomes an improper fraction, which contains identity-based terms. 
This in turn makes the resulting solution \eqref{ghost_brane} contain some identity-based terms.
Therefore,
we need careful treatment of these formal solutions.

\medskip
In the following,
we classify this class of solutions with respect to boundary states
and we propose 
a concrete example of 
the ghost brane solution.

\subsection{Boundary states}
We start from the calculation of
the closed string state $|B_\ast(\Psi)\rangle$
for the formal solutions~(\ref{ghost_brane}). 
Using the general formula,
we obtain
\begin{align}
\nonumber
x&=\lim_{z\to0}\left[\frac{z}{1-F(z)^2}\left(
\frac{1}{2}F(z)^2+F(z)F^\prime(z)\right)
+\frac{z}{1-H(z)^2}\left(
\frac{1}{2}H(z)^2+H(z)H^\prime(z)\right)\right]\\
&=-\frac{1}{2}\lim_{z\to0}\left[\frac{z\,\big(1-F(z)^2\big)^\prime}{1-F(z)^2}
+\frac{z\,\big(1-H(z)^2\big)^\prime}{1-H(z)^2}\right]\,.
\end{align}
In the same way,
$y$ is evaluated as
\begin{equation}
y=\frac{1}{2}\lim_{z\to0}\left[\frac{z\,\big(1-F(z)^2\big)^\prime}{1-F(z)^2}
+\frac{z\,\big(1-H(z)^2\big)^\prime}{1-H(z)^2}\right]\,.
\end{equation}
We notice that
the function $1-F(z)^2$
determines the property of the formal solution.
Let us consider the following two cases.
\begin{itemize}
\item $F(0)\neq1$, and therefore, $H(z)\neq1$

The regularity conditions imply
$x=y=0$,
and therefore,
$|B_*(\Psi)\rangle=|B\rangle$. 
The boundary state for the perturbative vacuum
can be reproduced
by this class of formal solutions.
However, they cannot written in the pure-gauge form~(\ref{pure_gauge_ansatz}),
and they seem not to be gauge equivalent to the perturbative vacuum solution.

\item $F(0)=1$, and therefore, $H(0)=1$

The regularity conditions imply
$1-F(z)^2= a z+\mathcal{O}(z^2)$
and $1-H(z)^2= -az+\mathcal{O}(z^2)$,
where $a$ is a non-zero constant.
We then obtain $x=-1$ and 
$y=1$.
Therefore,
$|B_*(\Psi)\rangle=-|B\rangle$,
which is nothing but the boundary state for the ghost D-brane.
\end{itemize}
Therefore, we conclude that
the following boundary states can be reproduced by
the formal solutions of the form \eqref{ghost_brane}: $|B_\ast(\Psi)\rangle=$ $|B\rangle$ and $-|B\rangle$.

\subsection{A candidate for the ghost brane solution}
In the last subsection
we showed that
the formal solution (\ref{ghost_brane})
reproduces the boundary state~$-|B\rangle$
when $1-F(z)^2=az+\mathcal{O}(z^2)$.
In this subsection we consider a particular choice of the function $F(z)$. 
For simplicity, let $F(K)^2$ be a meromorphic function of $K$,
and we set
\begin{align}
F(K)=\sqrt{\frac{1-pK}{1-qK}}
=\int_0^\infty dt_1\int_0^\infty dt_2 \, \frac{e^{-t_1}}{\sqrt{\pi t_1}}\frac{e^{-t_2}}{\sqrt{\pi t_2}} \, e^{(qt_1+pt_2)K}(1-pK)
\end{align}
to obtain\footnote{
Each term in~(\ref{real_ghost})
takes a similar form as
the solution for the tachyon vacuum
discussed in~\cite{Zeze:2010sr,Arroyo:2010sy}.}
\begin{equation}
\label{real_ghost}
\Psi_{\it{ghost}}
=\sqrt{\frac{1-pK}{1-qK}}c\frac{1-qK}{p-q}Bc\sqrt{\frac{1-pK}{1-qK}}+
\sqrt{\frac{1-qK}{1-pK}}c\frac{1-pK}{q-p}Bc\sqrt{\frac{1-qK}{1-pK}}\,,
\end{equation}
where
$p$ and $q$ are distinct positive constants. 
As we mentioned,
the formal solutions~(\ref{real_ghost}) contain some identity based terms.
They become more explicit
when we introduce a non-real solution~$\widetilde{\Psi}_{\it{ghost}}$
\begin{align}
\nonumber
\widetilde{\Psi}_{\it{ghost}}
&=\sqrt{\frac{1-pK}{1-qK}}\Psi_{\it{ghost}}\sqrt{\frac{1-qK}{1-pK}}\\
&=\frac{1-pK}{1-qK}c\frac{1-qK}{p-q}Bc+c\frac{1-pK}{q-p}Bc\frac{1-qK}{1-pK}\,,
\label{ghost_masuda}
\end{align}
which can be written in the following form:
\begin{equation}
\widetilde{\Psi}_{\it{ghost}}
=\frac{1}{q}\frac{1}{1-qK}c(qK-1)Bc+\frac{1}{p}c(pK-1)Bc\frac{1}{1-pK}+\frac{p+q}{pq}c-cKBc\,.
\label{ghost_identity}
\end{equation}
Here the first two terms are
some variants of
the simple solutions for tachyon condensation,
and the last two terms are identity-based terms. 
(It seems that, for any choice of $F(K)$, ghost brane solutions of the form \eqref{ghost_brane} always contain identity-based term as long as we assume our regularity conditions.)
Since the expression~(\ref{ghost_identity})
contains identity-based terms,
we need careful treatment
when we discuss the equation of motion
and the energy density of the solution. 

\paragraph{-- Energy }
Let us calculate the energy of the solution \eqref{real_ghost}.
We use the expression \eqref{ghost_masuda}\footnote{
We note that $\Psi$ and $\widetilde\Psi$ have the same energy density
because $\langle  \Psi\, Q \Psi\rangle=\langle \widetilde \Psi\, Q\widetilde \Psi\rangle$
and $\langle \Psi^3\rangle=\langle \widetilde\Psi^3\rangle$.} and
we define
\begin{align}
\widetilde \Psi_1=\frac{1-pK}{1-qK}c\frac{1-qK}{p-q}Bc\,,
\quad
\widetilde \Psi_2=
c\frac{1-pK}{q-p}Bc\frac{1-qK}{1-pK}\,.
\end{align}
Then, we can express the energy as a superposition of the correlation functions of wedge-based states as follows:
\begin{align}
\no
&\langle \widetilde \Psi_1\, Q\widetilde \Psi_1\rangle \\
\no
=\,\,&\frac{1}{q^2(p-q)^2}\int_0^\infty dx\int_0^\infty dy
\,e^{-(x+y)/q}\\
&\times(1-p\partial_x)(1-p\partial_y)(1-q\partial_u)(1-q\partial_v)\,\langle e^{Kx}ce^{Ku}Bc\, Q(e^{Ky}ce^{Kv}Bc)\rangle\Big|_{u\rightarrow 0,\,v\rightarrow 0}\\
\no
=\,\,&\frac{1}{q^2(p-q)^2}\int_0^\infty dx\int_0^\infty dy
\,e^{-(x+y)/q}\\
\no
&\times
\Bigg[\frac{\left(4 \pi ^2 p^2 x y+(x+y)^2 \left(2
   (p-x-y)^2-(x+y)^2\right)\right) \cos \left(\frac{2 \pi 
   x}{x+y}\right)}{2 \pi ^2 (x+y)^2}\\
\label{integrand}
&-\frac{2 (p-x-y)^2-(x+y)^2}{2 \pi
   ^2}-\frac{p (x-y) (-p+x+y) \sin \left(\frac{2 \pi 
   x}{x+y}\right)}{\pi  (x+y)}\Bigg]
 \\
=\,\,&-\frac{3}{\pi^2}\,,
\end{align}
where the expression of the integration kernel, $\langle e^{Kx}ce^{Ku}Bc\, Q(e^{Ky}ce^{Kv}Bc)\rangle$, is presented in \cite{Masuda}. 
Note that there exist no terms proportional to $\delta(x)\delta(y)$ in the integrand of (\ref{integrand}), which are contribution from the identity-based terms. 
Similarly, we obtain
\begin{align}
\langle \widetilde \Psi_2\, Q\widetilde \Psi_2\rangle&=-\frac{3}{\pi^2}\,,\\
\label{cross_term}
\langle \widetilde \Psi_1\, Q\widetilde \Psi_2\rangle&=0\,.
\end{align}
Since
correlation functions on a cylinder of
zero circumference have some ambiguity
in general,
we also confirm the following limit:
\begin{equation}
\lim_{\epsilon_1=a\epsilon_2 \to 0} \langle \widetilde \Psi_1e^{\epsilon_1 K}\, Q\widetilde \Psi_2e^{\epsilon_2 K}\rangle=0 \qquad \text{for }\quad ^\forall a>0\,,
\end{equation}
which reproduces the result in \eqref{cross_term}.
Therefore,
we conclude that
we can calculate the energy of the solution (\ref{real_ghost})
without ambiguity
unlike the identity based solutions\footnote{We should
note that
other regular solutions
such as
the simple solution for tachyon condensation~\cite{Erler:2009uj}
can be written in the form containing identity-based terms:
it can be written as
\begin{equation}
\Psi_{\it{ES}}=\frac{1}{1-K}c(K-1)Bc=\frac{1}{1-K}Bc(1-K)c-c\,.
\end{equation}
},
and the resulting energy density is
\begin{equation}
E=-\frac{1}{\pi^2}\;,
\end{equation}
which coincides with the energy density of the ghost D-brane~\cite{Okuda:2006fb}.

\paragraph{-- Properties }
We summarize our current understanding
of our ghost brane solution below:
\begin{itemize}
\item Its component fields are all finite:
$\langle\varphi,\Psi_{\it{ghost}}\rangle={\rm finite}$.
\item It satisfies the equation of motion
when contracted to the state in the Fock space
and the solution itself:
$\langle\varphi,Q\Psi_{\it{ghost}}+\Psi_{\it{ghost}}^2\rangle=0$
and
$\langle\Psi_{\it{ghost}},Q\Psi_{\it{ghost}}+\Psi_{\it{ghost}}^2\rangle=0$.
 \item Its energy density is twice as that of the tachyon vacuum solution: $\displaystyle{E = - \frac{1}{\pi^2}}$,
which coincides with that of the ghost D-brane~\cite{Okuda:2006fb}.
\item It reproduces the boundary state for the ghost D-brane:
$|B_\ast(\Psi_{\it{ghost}})\rangle=-|B\rangle$.
\item It satisfies the so-called
weak consistency condition introduced in~\cite{Erler:2012qn}.\footnote{
See appendix~\ref{app_pro} for details.}

\end{itemize}
At this point, it is not clear whether the solution can be regarded as a physical solution or not.
All we can say is that the properties itemized above seem not to prohibit it. 
Further investigation would be required. 

\section{Discussion}
\label{discussion_sec}
\setcounter{equation}{0}
In this paper
we evaluated boundary states
for general string fields in the $KBc$ subalgebra
under some regularity conditions.
By requiring the $s$-independence
as a necessary condition
to satisfy the equation of motion,
we showed that
there are only three possible boundary states
in our class:
$|B_*\rangle = \pm|B\rangle$ and $0$.
Our results seem to suggest
that we are not able to
construct multiple D-brane solutions
in the $KBc$ subalgebra
without relaxing our regularity conditions.
However,
as we discussed in section~\ref{subsection_s-dep},
even if we consider singular string fields
such as proposed multiple D-brane solutions,
the expected boundary states
cannot be reproduced
as long as we express them as a limit of
regular string fields and
formally use our formula (\ref{def_alpha}),
(\ref{def_beta}),
and (\ref{B_ast}).

\medskip
To be precise,
our 
derivation 
was based on the Schnabl gauge calculation.
As discussed in~\cite{Kiermaier:2007jg},
Schnabl gauge can be understood
as a singular limit of a one parameter family of
regular linear $b$ gauges called $\lambda$-regularized gauges.
Originally in~\cite{Kiermaier:2008qu},
the Schnabl gauge calculation was introduced
based on the discussion~\cite{Kiermaier:2008jy} using $\lambda$-regularized gauges.
Although the calculation seems to be valid
for regular string fields,
it is not clear whether it is applicable
for singular string fields.
It requires 
careful investigations
using $\lambda$-regularized gauges
in order to clarify this point
and complete our discussion.
Some preliminary works in this direction
are underway and
we hope to report our progress
elsewhere.

\bigskip
\noindent
{\bf \large Acknowledgments}

We are grateful to
Yuji Okawa for helpful discussion
and careful reading of the manuscript.
We would like to thank Mitsuhiro Kato and Masaki Murata for
valuable discussion.
We also thank
Hashimoto Mathematical Physics Laboratory at RIKEN
for providing us
with a stimulating environment for discussion in
the informal mini workshop on string field theory. 
We thank the Yukawa Institute for Theoretical Physics at Kyoto University. Discussions during the YITP workshop on ``Field Theory and String Theory'' (YITP-W-12-05) were useful to complete this work. 
We also thank Hiroyuki Hata and Toshiko Kojita for
their kind help during this workshop. 
The work of T.N. was supported in part by JSPS Grant-in-Aid for JSPS Fellows.

\bigskip
\noindent
{\bf \large Note added}

The contents of appendix B.2 are added in the replacement from v1 to v2
on arXiv.
While we were preparing the replacement,
the paper \cite{Erler:2012he} appeared on arXiv,
in which a level expansion method in the sliver frame was proposed
and the homomorphism of the $KBc$ subalgebra in appendix B.2 was also pointed
out in that context. 
We would like to thank Ted Erler for correspondence. 
\appendix
\section{Projectors and consistency conditions}
\setcounter{equation}{0}
\label{app_pro}
In~\cite{Erler:2012qn},
extending Ellwood's discussion~\cite{Ellwood:2009zf}
on singular gauge transformations in open string field theory,
Erler and Maccaferri showed that
there is always a nonzero left gauge transformation~$U_{\widetilde{\Psi}\Psi}$
connecting any pair of classical solutions $\Psi$ and $\widetilde{\Psi}$:
\begin{equation}
\label{left-gauge}
QU_{\widetilde{\Psi}\Psi}+\widetilde{\Psi}U_{\widetilde{\Psi}\Psi}=U_{\widetilde{\Psi}\Psi}\Psi\,.
\end{equation}
When $\Psi$ and $\widetilde{\Psi}$ describe different backgrounds,
$U_{\widetilde{\Psi}\Psi}$ has a non-trivial kernel,
and it is expected that
the kernel
 of $U_{\widetilde{\Psi}\Psi}$ captures some properties of the solutions.
They
introduced a projector $X^\infty_{\widetilde{\Psi}\Psi}$ onto the kernel
and carefully investigated its property.
They found that the projectors for known solutions
have interesting structures associated with
the BCFT described by the solutions
and they called the projector $X^\infty_{\widetilde{\Psi}\Psi}$
the boundary condition changing projector.
Although its property is not fully understood yet,
we expect that it
would be helpful for
constructing new solutions.\footnote{
See~\cite{Erler:2012qr} for an interesting application of the projector
to determining the so-called phantom term.}
In this appendix
we first investigate
the structure of the projector
in the $KBc$ subalgebra,
and then,
we apply
the
result to the ghost brane solution~(\ref{real_ghost}).
We also show that the ghost brane solution
satisfy the so-called week consistency condition
introduced in~\cite{Erler:2012qn}.

\subsection{Projectors in $KBc$ subalgebra}
As discussed in~\cite{Erler:2012qn},
for any solutions $\Psi$ and $\widetilde{\Psi}$,
we can find a non-zero left gauge transformation
$U_{\widetilde{\Psi}\Psi}$
in the following way:
\begin{equation}
\label{def_U_12}
U_{\widetilde{\Psi}\Psi}=Q\,\textbf{b}+\widetilde{\Psi} \textbf{b}+\textbf{b}\Psi\,,
\end{equation}
where $\textbf{b}$ is a Grassmann-odd state of ghost number $-1$.
The relation~(\ref{left-gauge})
follows from the equations of motion for $\Psi$ and $\widetilde{\Psi}$.
In this paper
we define the projector $X_{\widetilde{\Psi}\Psi}^\infty$
onto the kernel of $U_{\widetilde{\Psi}\Psi}$ by\footnote{
We assume that we can take a small number $\epsilon$
so that $\epsilon+U$ does not have any non-trivial kernel.
For example, we take $\epsilon$ to be a negative real number
when $U=K$.
}
\begin{equation}
\label{def_projector}
X_{\widetilde{\Psi}\Psi}^\infty=\lim_{\epsilon\to0}\frac{\epsilon}{\epsilon+U_{\widetilde{\Psi}\Psi}}\,.
\end{equation}
In this subsection
we calculate $U_{\widetilde{\Psi}\Psi}$ defined by~(\ref{def_U_12})
and the associated projector $X^\infty_{\widetilde{\Psi}\Psi}$
for general string fields in the $KBc$ subalgebra\footnote{
In the calculation of this appendix,
we assume that $F_i(K)$, $G_i(K)$, $H_i(K)$,
$\widetilde{F}_i(K)$, $\widetilde{G}_i(K)$, and $\widetilde{H}_i(K)$
satisfy the regularity condition I introduced in~(\ref{regularity}).
}:
\begin{equation}
\label{general_Psi_pro}
\Psi=\sum_iF_i(K) cB G_i(K) c H_i(K)\,,
\quad
\widetilde{\Psi}=\sum_i\widetilde{F}_i(K)cB\widetilde{G}_i(K)c\widetilde{H}_i(K)\,.
\end{equation}
We note that $U_{\widetilde{\Psi}\Psi}$ and $X^\infty_{\widetilde{\Psi}\Psi}$ are well-defined
by~(\ref{def_U_12}) and~(\ref{def_projector})
although $\Psi$ and $\widetilde{\Psi}$ in (\ref{general_Psi_pro})
do not necessarily satisfy the equation of motion
and $U_{\widetilde{\Psi}\Psi}$ do not necessarily
reproduce the left gauge transformation~(\ref{left-gauge}).

\subsubsection{General form of the projector}
In the $KBc$ subalgebra,
the general form of the state $\textbf{b}$ of ghost number $-1$
is given by
\begin{equation}
\label{b_in_KBc}
\textbf{b}=BM(K)\,,
\end{equation}
where $M(K)$ is a function of $K$
satisfying the regularity condition I 
and we also assume that
it does not have a non-trivial kernel.
Substituting this general form
into the definition~(\ref{def_U_12}),
we obtain
\begin{align}
\nonumber
U_{\widetilde{\Psi}\Psi}&=KM+\sum_i\widetilde{F}_icB\widetilde{G}_i\widetilde{H}_iM+M\sum_iF_iG_iBcH_i\\
&=(K+\sum_i\widetilde{F}_i\widetilde{G}_i\widetilde{H}_i)M-\sum_i\widetilde{F}_iBc\widetilde{G}_i\widetilde{H}_iM+M\sum_iF_iG_iBcH_i\,.\label{general_gauge}
\end{align}
It is convenient to introduce the following formula
in the $KBc$ subalgebra:
\begin{equation}
(H+\sum_iI_iBcJ_i)^{-1}=\frac{1}{H+\sum_{i}I_iJ_i}(1+\sum_jI_jcBJ_jH^{-1})\,,
\end{equation}
where $H$, $I_i$, and $J_i$ are functions of $K$
satisfying $H\neq0$ and $H+\sum_iI_iJ_i\neq0$.
Using this formula,
we calculate $(\epsilon+U_{\widetilde{\Psi}\Psi})^{-1}$ as
\begin{align}
\nonumber
(\epsilon+U_{\widetilde{\Psi}\Psi})^{-1}&=\frac{1}{\epsilon+(K+\sum_iF_iG_iH_i)M}\\
\nonumber
&\,\,\,\,-\frac{1}{\epsilon+(K+\sum_iF_iG_iH_i)M}\Big(\sum_j\widetilde{F}_jcB\widetilde{G}_j\widetilde{H}_jM\Big)\frac{1}{\epsilon+(K+\sum_k\widetilde{F}_k\widetilde{G}_k\widetilde{H}_k)M}\\
&\,\,\,\,+\frac{1}{\epsilon+(K+\sum_iF_iG_iH_i)M}\Big(M\sum_jF_jG_jcBH_j\Big)\frac{1}{\epsilon+(K+\sum_k\widetilde{F}_k\widetilde{G}_k\widetilde{H}_k)M}\,.
\end{align}
Then,
the associated projector $X^\infty_{\widetilde{\Psi}\Psi}$ is given by
\begin{equation}
\label{general_projector}
X^\infty_{\widetilde{\Psi}\Psi}
=\lim_{\epsilon\to0}\frac{\epsilon}{\epsilon+U_{\widetilde{\Psi}\Psi}}
=\mathcal{P}_{\Psi}+\mathcal{P}^{cB}_{\Psi\widetilde{\Psi}}\,,
\end{equation}
where
\begin{align}
\label{P_psi}
\mathcal{P}_{\Psi}&=\lim_{\epsilon\to0}\frac{\epsilon}{\epsilon+(K+\sum_iF_iG_iH_i)M}\,,\\
\nonumber
\mathcal{P}^{cB}_{\Psi\widetilde{\Psi}}&=
\lim_{\epsilon\to0}\frac{\epsilon}{\epsilon+(K+\sum_iF_iG_iH_i)M}\Big(M\sum_jF_jG_jcBH_j\Big)\frac{1}{\epsilon+(K+\sum_k\widetilde{F}_k\widetilde{G}_k\widetilde{H}_k)M}\\
&-\lim_{\epsilon\to0}\frac{\epsilon}{\epsilon+(K+\sum_iF_iG_iH_i)M}\Big(\sum_j\widetilde{F}_jcB\widetilde{G}_j\widetilde{H}_jM\Big)\frac{1}{\epsilon+(K+\sum_k\widetilde{F}_k\widetilde{G}_k\widetilde{H}_k)M}\,.\,
\end{align}
We notice that
the property of the projector is determined
by $\sum_iF_iG_iH_i$ and $\sum_i\widetilde{F}_i\widetilde{G}_i\widetilde{H}_i$.

\subsubsection{From tachyon vacuum to general string fields}
Let us consider the case when
$\widetilde{\Psi}$ is a solution for the tachyon vacuum:
\begin{equation}
\widetilde{\Psi}=\Psi_{\rm tv}=I(K)c\frac{KB}{1-I(K)^2}cI(K)\,,
\end{equation}
where $\displaystyle{I(K)=e^{K/2}}$ for Schnabl's solution~\cite{Schnabl:2005gv} 
and $\displaystyle{I(K)=\frac{1}{\sqrt{1-K}}}$ for the simple solution~\cite{Erler:2009uj}.
In this case,
$\mathcal{P}^{cB}_{\Psi\widetilde{\Psi}}$ is given by
\begin{equation}
\label{X_t_to_g}
\mathcal{P}^{cB}_{\Psi\widetilde{\Psi}}=\lim_{\epsilon\to0}\frac{\epsilon}{\epsilon+M(K+\sum_iF_iG_iH_i)}\Big(M\sum_iF_iG_icBH_i-IcB\frac{K}{1-I^2}IM\Big)\frac{1}{\frac{K}{1-I^2}M+\epsilon}\,.
\end{equation}
We notice that
we can take the limit
$\lim_{\epsilon\to0}(\frac{K}{1-I^2}M+\epsilon)^{-1}=\frac{1-I^2}{K}M^{-1}$
without using singular expression
because $\frac{K}{1-I^2}M$ does not have a non-trivial kernel.
Therefore,
we can rewrite the projector $X^\infty_{\Psi_{\rm tv}\Psi}$
as
\begin{equation}
X^\infty_{\Psi_{\rm tv}\Psi}=\mathcal{P}_\Psi+\mathcal{P}_\Psi M\sum_iF_iG_icBH_i\frac{1-I^2}{KM}
-\mathcal{P}_\Psi IcBI\,.\label{projector_tv}
\end{equation}

\subsubsection{From perturbative vacuum to general string fields}
In the case when
$\widetilde{\Psi}$ is a solution for the perturbative vacuum
\begin{equation}
\widetilde{\Psi}=\Psi_{\rm pv}=0\,,
\end{equation}
the projector $X^\infty_{\Psi_{\rm pv}\Psi}$
is given by
\begin{align}
\label{X_p_to_g}
X^\infty_{\Psi_{\rm pv}\Psi}
=\lim_{\epsilon\to0}\frac{\epsilon}{\epsilon+M(K+\sum_iF_iG_iH_i)}
+\lim_{\epsilon\to0}\frac{M}{\epsilon+M(K+\sum_jF_jG_jH_j)}\sum_iF_iG_icBH_i\frac{\epsilon}{\epsilon+KM}\,. 
\end{align}
Unlike the previous case,
the limit $\epsilon\to0$ can be singular in general
because the state $K$ has a non-trivial kernel.

\subsection{Consistency condition for the ghost brane solution}
In this subsection
we apply our result in the previous subsection
to the ghost brane solution~(\ref{real_ghost}),
and we show that it satisfies the so-called
weak consistency condition~\cite{Erler:2012qn}.

\subsubsection{Consistency conditions}
When two solutions
$\Psi$ and $\widetilde{\Psi}$ are connected by
a left gauge transformation $U_{\tilde\Psi\Psi}$,
it follows 
from the expression (\ref{left-gauge})
that
\begin{align}
	\mathrm{Im}\;(Q+\widetilde\Psi)U_{\tilde\Psi\Psi} \subseteq \mathrm{Im}\;U_{\tilde\Psi\Psi}\,,
\end{align}
which is called the strong consistency condition~\cite{Erler:2012qn}.
Provided that the projector $X^\infty_{\tilde\Psi\Psi}$
onto the kernel of $U_{\tilde\Psi\Psi}$ exists,
$X^\infty_{\tilde\Psi\Psi}$ satisfies
\begin{align}
	\mathrm{ker}\;U_{\tilde\Psi\Psi} = \mathrm{Im}\;X^\infty_{\tilde\Psi\Psi},
	\quad \mathrm{Im}\;U_{\tilde\Psi\Psi} \subseteq \mathrm{ker}\;X^\infty_{\tilde\Psi\Psi}\;,
\end{align}
and therefore,
the following relation is implied:
\begin{align}
\mathrm{Im}\;(Q+\widetilde\Psi)U_{\tilde\Psi\Psi} \subseteq \mathrm{Im}\;U_{\tilde\Psi\Psi} \subseteq \mathrm{ker}\;X^\infty_{\tilde\Psi\Psi}\;.
\end{align}
In other words, 
\begin{align}
X^\infty_{\tilde\Psi\Psi} Q_{\tilde\Psi}U_{\tilde\Psi\Psi} = 0\,,
\label{weak_consist}
\end{align}
which is called the weak consistency condition~\cite{Erler:2012qn}.
The relation~(\ref{weak_consist})
should hold
if two solutions are connected by left gauge transformation and the projector $X^\infty_{\tilde\Psi\Psi}$ exists.
In what follows,
we show that the ghost brane solution~(\ref{real_ghost})
satisfies the weak consistency condition~(\ref{weak_consist})
when connected to the tachyon vacuum and the perturbative vacuum.

\subsubsection{From tachyon vacuum to the ghost brane}
Let us start from the case of the tachyon vacuum.
Using (\ref{general_gauge}) and (\ref{projector_tv}),
the gauge transformation $U_{\Psi_{\mathrm{tv}}\Psi_{\mathrm{gh}}}$
and 
the projector $X_{\Psi_{\mathrm{tv}}\Psi_{\mathrm{gh}}}^\infty$
are given by
\begin{align}
U_{\Psi_{\mathrm{tv}}\Psi_{\mathrm{gh}}}
&= \frac{K}{1-I^2}M - IBc\frac{K}{1-I^2}IM + MF\frac{K}{1-F^2}BcF(1-H^2)\,,\\
X_{\Psi_{\mathrm{tv}}\Psi_{\mathrm{gh}}}^\infty
&= 1 + MF\frac{K}{1-F^2}cBF(1-H^2)\frac{1-I^2}{KM} - IcBI\,.
\end{align}
We then obtain that
\begin{align}
QU_{{\Psi_\mathrm{tv}}\Psi_{\mathrm{gh}}} 
  + \Psi_{\mathrm{tv}}U_{{\Psi_\mathrm{tv}}\Psi_{\mathrm{gh}}} 
= MF\frac{K}{1-F^2}cBKcF(1-H^2) + Ic\frac{K}{1-I^2}IMF\frac{K}{1-F^2}BcF(1-H^2) \,.
\end{align}
It is straightforward to show that
\begin{align}
X^\infty_{\Psi_{\mathrm{tv}}\Psi_{\mathrm{gh}}} Q_{\Psi_\mathrm{tv}}U_{{\Psi_\mathrm{tv}}\Psi_{\mathrm{gh}}}=0\,,
\end{align}
and therefore,
the ghost brane solution satisfies the weak consistency condition
when connected to the tachyon vacuum. 

\subsubsection{From perturbative vacuum to the ghost brane}
We then consider the case of the perturbative vacuum.
Using (\ref{general_gauge}) and (\ref{projector_tv}),
the gauge transformation $U_{\Psi_{\mathrm{tv}}\Psi_{\mathrm{gh}}}$
and 
the projector $X_{\Psi_{\mathrm{tv}}\Psi_{\mathrm{gh}}}^\infty$
are given by\footnote{
We use the symbol $\widetilde \Omega^{\infty}$
to denote sliver-like states as in section \ref{subsubsub_phantom}.}
\begin{align}
	U_{\Psi_{\mathrm{pv}}\Psi_{\mathrm{gh}}}
&= KM + M\frac{K}{1-F^2}BcF(1-H^2)\,,\\
	X^\infty_{\Psi_{\mathrm{pv}}\Psi_{\mathrm{gh}}}
	&=1 + \lim_{\epsilon \rightarrow 0}MF\frac{K}{1-F^2}cBF(1-H^2)\frac{1}{\epsilon + KM}\no\\
	&=1 + MF\frac{K}{1-F^2}cBF\frac{1-H^2}{K}KM^{-1}(1-\widetilde \Omega^\infty)\,,
\end{align}
where we used
\begin{align}
\lim_{\epsilon \rightarrow 0}(1-H^2)\frac{1}{\epsilon + KM}
=\frac{1-H^2}{K}M^{-1}
\lim_{\epsilon \rightarrow 0}\frac{KM}{\epsilon + KM}
=\frac{1-H^2}{K}M^{-1}(1-\widetilde \Omega^\infty)\,.
\end{align}
We then obtain that
\begin{align}
QU_{\Psi_{\mathrm{pv}}\Psi_{\mathrm{gh}}} = MF\frac{K}{1-F^2}cBKcF(1-H^2)\,.
\end{align}
It is straightforward to show that
\begin{align}
&   X_{\Psi_{\mathrm{pv}}\Psi_{\mathrm{gh}}}^\infty QU_{\Psi_{\mathrm{pv}}\Psi_{\mathrm{gh}}} \no\\
& = MF\frac{K}{1-F^2}cBKcF(1-H^2) \no\\
&\quad  + MF\frac{K}{1-F^2}cBF\frac{1-H^2}{K}M^{-1}(1-\widetilde \Omega^\infty)MF\frac{K}{1-F^2}cBKcF(1-H^2) \no\\
& = MF\frac{K}{1-F^2}cBKcF(1-H^2) - MF\frac{K}{1-F^2}cBKcF(1-H^2)\no\\
& = 0\,,
\end{align}
where we used $FH=1$ and 
formally used 
$\Omega^\infty K =0$.
Therefore,
the ghost brane solution satisfies the weak consistency condition
when connected to the perturbative vacuum.\footnote{
In~\cite{Erler:2012qn},
it was shown that
another candidate for the ghost brane solution,
which was
constructed in a similar way as the proposed multiple D-brane solutions~\cite{Murata:2011ex,Murata:2011ep},
does not satisfy the weak consistency condition
when connected to the 
tachyon vacuum.
However,
our solution is constructed in a different way,
and it satisfies
both the equation of motion and
the weak consistency condition.
}

\section{Algebraic structure}  
\label{appendix_B}
\setcounter{equation}{0}
The results obtained in this paper can be regarded as 
classification of classical solutions in the 
$KBc$ subalgebra with respect to corresponding boundary states.
On the other hand, works by Erler \cite{Erler:2006ww} 
and Murata and Schnabl \cite{Murata:2011ex, Murata:2011ep} 
are classification with respect to the energy.
In this short appendix, 
we attempt to clarify the underlying algebraic structure 
in these discussions from the viewpoint of gauge structures.  
We also provide
a proof that the solution in the class \eqref{ghost_brane} cannot be
written in the pure-gauge form~(\ref{pure_gauge_ansatz}).

\subsection{On the relation between \eqref{pure_gauge_ansatz} and \eqref{ghost_brane}}
Let $\Psi$ be a solution to the equation of motion. 
Define the gauge parameter $U$ by\footnote{As explained 
in section~\ref{pure_geuge_ansatz}, we do not 
assume gauge parameters
$U$ and $U^{-1}$ are regular.
} 
\begin{equation}\label{U1APsi}
U=1+A\Psi\,,
\end{equation}
where $A$ denotes the formal homotopy operator 
around the perturbative vacuum,
$A={B}/{K}$\,.
Supposing that the inverse of $U$ exists,
we can rewrite $\Psi$ as follows:\footnote
{This procedure to render solutions into 
pure-gauge form is essentially the same as 
that presented by Ellwood in section 2.1 of \cite{Ellwood:2009zf}.
}
\begin{equation}\label{PsiU-1QU}
\Psi=U^{-1}QU\,.
\end{equation}
Note that
the normalization of  the gauge parameter $U$
is different from that in section 4.
Then consider when the inverse of $U$ exist. 
The answer is already presented in section 4; suppose $U$ is expressed as
\begin{equation}\label{U}
U=1+\sum_iv_i(K)Bcw_i(K)\,.
\end{equation}
If $1+\sum_i v_i w_i\ne 0\,$, we can find its formal inverse\footnote
{When $1+\sum_i v_iw_i=0\,$, the string field $U$ takes the following form: 
\begin{equation}\notag
U=-\sum_i v_i(K) cB w_i(K)\,.
\end{equation}
We do not consider its inverse, 
just as we do not consider the inverse of $cB$. },
\begin{equation}
U^{-1}=1-\frac{1}{1+\sum_iv_i w_i}\sum_iv_i Bc{w_i}\,.
\end{equation}
The condition $1+\sum_i v_i w_i\ne 0$ can be also expressed as
\begin{equation}
UB\ne 0\,.
\end{equation}
Therefore, $\Psi$ can be written in the pure-gauge form \eqref{pure_gauge_ansatz} if
\begin{equation}\label{cond.pg}
(1+A\Psi)B\ne 0\,.
\end{equation}

Next, we prove the inverse statement: if $\Psi$ can be written in 
the pure-gauge form \eqref{pure_gauge_ansatz},
then $\Psi$ satisfies the equation \eqref{cond.pg}.
We can directly check the statement using the explicit expression \eqref{pure_gauge_ansatz}.

\medskip
To summarize,
$\Psi$ can be written in the pure-gauge form \eqref{pure_gauge_ansatz} if and only if
\begin{equation}
(1+A\Psi)B\ne 0\,.
\end{equation}

\medskip
Then, let us apply above discussion to the formal solution \eqref{ghost_brane}.
Since \eqref{ghost_brane} satisfies $(1+A\Psi)B= 0\,$,
it can not be written in the pure-gauge form \eqref{pure_gauge_ansatz}.
We note that since $(1+A\Psi)B \neq 0$ for (\ref{pure_another}),
it can be written in the pure-gauge form.

\paragraph{-- Discriminant function $d_\Psi(x)$}
Let us rephrase above discussion in terms of
the discriminant function $d_\Psi(x)$ defined by
\begin{equation}
d_\Psi(K)B=\left(1+A\Psi\right)B\,. 
\end{equation}
The solution $\Psi$ can be written in the form \eqref{pure_gauge_ansatz} if and only if
$d_\Psi(K)\ne 0$.
It is also true that $\Psi$ is pure-gauge if and only if $d_\Psi(0)\ne 0,\, \infty$.

\medskip
We notice that
the discriminant function $d_\Psi(x)$ was used
to classify formal pure-gauge solutions with respect to energy or boundary states.
For Okawa's formal solutions \eqref{Okawa's_solution}, the discriminant function is  
\begin{equation}
d_\Psi(x)=\frac{1}{1-F(x)^2}\,,
\end{equation}
which was used in \cite{Murata:2011ex,Murata:2011ep} to classify 
the formal solutions by energy. 
For the formal solution \eqref{pure_gauge_ansatz},  
the discriminant function is given by
\begin{equation}
d_\Psi(x)=\frac{1}{1-\sum_{i}I_i(x)J_j(x)}\,,
\end{equation}
which we used in section \ref{pure_sec}
to classify the formal solutions by boundary states. 
Indeed, both of these classifications are determined by the degree of poles or zeros of $d_\Psi(x)$ at $x=0\,$.  

\subsection{On the classification of the formal pure-gauge solutions}
\label{B.2}

In this subsection, we consider a group structure of ghost number zero string fields in 
the $KBc$ subalgebra, which leads us to a classification of the formal pure-gauge solutions
 in terms of discriminant function. 
This classification may be regarded as an simple generalization of that 
given by Murata and Schnabl \cite{Murata:2011ex, Murata:2011ep}. 
It is also compatible with the discussion in section 4.2, which is classification of 
solutions under the regularity condition presented in section 2. 

\paragraph{-- A class of gauge parameters $S_0$} 
To begin with, we define a function $d_U(x)$, 
which is frequently used throughout the present subsection. 
Let $U$ be a ghost number zero element in the $KBc$ subalgebra. 
We define a function $d_U(x)$ for $U$
as follows:
\begin{equation}\label{dU}
d_U(K)B=UB\,.
\end{equation}
The function $d_U$ preserves multiplication of $U$'s:
\begin{equation}\label{dU1U2}
d_{U_1 U_2}=d_{U_1}d_{U_2}\,.
\end{equation}
Using the function $d_U$, we define a class of gauge parameters $S^\prime_0$ as 
\begin{equation}
S^\prime_0=\Big\{\,U=1+\sum v_i Bc w_i \,{\Big |}\, d_U(x) \ne 0 \,\Big \}\,,
\end{equation}
which is closed under star multiplication and inverses. 
It is important that \eqref{U1APsi} and \eqref{PsiU-1QU} give 
one to one correspondence between the space of formal pure-gauge 
solutions \eqref{pure_gauge_ansatz} and gauge parameters $S_0^\prime$. 
In particular, any solution of the form 
\eqref{pure_gauge_ansatz} can be written as $\Psi=U^{-1}QU$, where $U\in S^\prime_0$. 
In the following, we always keep this one to one correspondence in mind. 
Note that the discriminant function $d_\Psi$ 
for $\Psi=U^{-1}QU$ $(U\in S^\prime_0)$ is given by $d_U$. 
For this reason, we also call $d_U(x)$ a discriminant function, as well as $d_\Psi(x)$.  

For later use, we also define a class of gauge parameters $S_0$ as follows:
\begin{equation}
S_0=\Big\{\,U \in S^\prime_0 \,{\Big |}\, 
d_U(x) \text{ is meromorphic around $x=0$.\,} \Big \}\,,
\end{equation}
which is also closed under multiplication and inverses. 

\paragraph{-- Multiplicity }
Since the discriminant function $d_U(x)$ for $U\in S_0$ is meromorphic around $x=0$, 
we can expand $d_U(x)$ as follows:
\begin{equation}\label{multiplicity_definition}
d_U(x)=1+\sum_i v_i(x)w_i(x)=\sum_{m=n}^\infty a_m x^m\,.
\end{equation}
We define the multiplicity\footnote{
We expect that our multiplicity seems to be same as the winding number 
discussed by Hata and Kojita \cite{Hata:2011ke}
up to some subtleties arising from regularization of solutions.}
 $n$ of a gauge parameter $U\in S_0$ by the integer $n$ in \eqref{multiplicity_definition}. 
If $U$ possesses multiplicity $m$ and $V$ possesses multiplicity $n$, 
then, from the property \eqref{dU1U2}, the product $UV$
 possesses multiplicity $m+n$. 
We also note that any two gauge parameters $U_1$ and $U_2$ with the same multiplicity 
can be connected by a multiplicity zero gauge parameter.  
The class of multiplicity zero elements $R$ make a normal subgroup of $S_0$:
\begin{equation}\label{R}
R=\big\{U\in S_0 \big| d_U(0)\ne 0,\ d_U(0)\ne \infty\big\}\,.
\end{equation} 
Since $R$ is a normal subgroup of $S_0$, we can consider the quotient group $S_0/R$, 
which is isomorphic to $\mathbb Z$. 

\paragraph{-- Classification of pure-gauge solutions by multiplicity}
Now, let us consider a class of formal pure-gauge solutions
whose discriminant functions $d_\Psi(x)$ are meromorphic around $x = 0$.
This class of formal solutions correspond to elements of $S_0$,
and we define the multiplicity of such formal solutions
by that of the corresponding gauge parameters $U \in S_0$. 
Any two solutions $\Psi_1$ and $\Psi_2$ of the same multiplicity can be connected 
by a multiplicity zero gauge parameter $U\in R$.

We note that if we impose our regularity condition on the formal pure-gauge solutions, 
 the classification by multiplicity reduces to the discussion in section 4.2. 
Similarly, if we concentrate on the Okawa type solutions, 
the classification by multiplicity reduces to the discussion 
given by Murata and Schnabl~\cite{Murata:2011ex, Murata:2011ep}. 

\paragraph{-- On the property of $R$}
The classification by the multiplicity is natural from the algebraic point of view.
Then, we would like to consider whether it has some physical meaning
or whether it is related to gauge equivalence of solutions.
Our question can be rephrased as whether the ``gauge" transformations corresponding to elements of $R$ possess some regular property or not,
and, in the following, we give a naive discussion on their regularity.  

Let $U$ be an element of $R$. $U$ can always be factorized into the following form:
\begin{equation}
U=U_1\ast U_2\ast\dots \ast U_n\,, 
\end{equation}
where each $U_i$ takes the form 
\begin{equation}
U_i=1+v_i(K)Bc\,w_i(K)\,.
\end{equation}
Functions $v_i$ and $w_i$ are meromorphic around the origin. 
Then, define $\tilde U_i$ by  
\begin{equation}
\begin{split}
\tilde U_i =w_i(K)\,U_i w_i(K)^{-1}=1+v_i(K)w_i(K)Bc\,.
\end{split}
\end{equation}
$U$ can be written as
\begin{equation}\label{B.19}
U=w_1^{-1}\,\tilde U_1 w_1 w_2^{-1}\,\tilde U_2 w_2\dots w_n^{-1}\,\tilde U_n w_n\,.
\end{equation}
The gauge transformation by $U$ is equivalent to 
a reciprocal sequence of gauge transformation by $\tilde U_i$ 
and similarity transformation $\Psi\rightarrow w_i^{-1}\Psi w_i$ 
or $\Psi\rightarrow w_i\Psi w_i^{-1}$. 
These similarity transformations are expected not to change energy of solutions. 
Then, {\bf if} we omit all the $w_i$'s in \eqref{B.19}, $U$ reduces to $\tilde U$, where
\begin{equation}
\tilde U=\tilde U_1\ast \tilde U_2\ast\dots \ast \tilde U_n\,. 
\end{equation}
It is not difficult to see that the gauge parameter $\tilde U$ is in the 
following form:
\begin{equation}
\tilde U=1+ u (K)Bc\,.
\end{equation}
Since $d_{U_i}=d_{\tilde U_i}$, it is also obvious that
 $d_{\tilde U}=0$ and ${\tilde U}\in R\,$. 
Then, $\tilde U$ and $\tilde U^{-1}$ do not contain 
singular string fields like $1/K$. 
We naturally expect that gauge transformation by $\tilde U$ is regular. 
We admit, however, that the above discussion 
is not sufficient to prove regularity of $U$. 

\paragraph{-- Homomorphism of the $KBc$ subalgebra}
In the last of this appendix,
we consider a series of formal homomorphisms $h_g$,
 which is a generalization of the first example given by Erler \cite{Erler:2010zza}. 
We define the action of $h_g$ by\footnote{This generalization is also pointed out in \cite{Erler:2012he} by Erler.}
\begin{eqnarray}
h_g(K) &=& g(K) \equiv \tilde K\,, \\
h_g(B) &=& {g(K)B}/{K} \equiv \tilde B\,,\\
h_g(c)&=& c({KB}/{g(K)})c\equiv \tilde c\,,
\end{eqnarray}
and
\begin{equation}
h_g(\Phi_1 \ast \Phi_2)=h_g(\Phi_1) \ast h_g(\Phi_2)\,.
\end{equation}
Here the subscript $g$ of $h_g$
stands for the function $g(K)$.
It is not difficult to confirm that $\tilde K$, $\tilde B$ and $\tilde c$ 
satisfy the same algebraic relations as $K$, $B$, and $c$:
\begin{equation}
\label{KBc_appB}
\tilde B^2=\tilde c^2=0\,,\quad \{\tilde B,\tilde c\}=1\,,\quad
Q\tilde B=\tilde K\,,\quad
Q\tilde c=\tilde c\tilde K\tilde c\,,\quad
[\tilde K,\tilde B]=0\,,
\end{equation}
When the function $g(x)$ is meromorphic around $x=0$,  
$h_g(S_0)$ is a subgroup of $S_0\,$. 
We remark that multiplicity of an element of $S_0$ changes as $n\rightarrow m n\,$, where  
\begin{equation}
m=\lim_{x\to 0}\frac{x\,d(\ln g(x))}{dx}\,.
\end{equation}

The action of $h_{g}$ on the formal pure-gauge solutions takes quite simple form. 
Let $\Psi_F$ be an Okawa type formal solution,
\begin{equation}
\Psi_{F}=F(K)c\frac{KB}{1-F(K)^2}cF(K)\,.
\end{equation} 
Then the action of $h_{g}$ is given by
\begin{equation}
h_g(\Psi_F)=\Psi_{\tilde F}\,,
\end{equation}
where 
\begin{equation}
\tilde F(K)=F(\tilde K)\,.
\end{equation}
Similarly, let $\Psi_{\{I_i,J_i\}}$ be a solution of the form (4.4).
Then the action of $h_g$ is
\begin{equation}
h_g(\Psi_{\{I_i,J_i\}})=\Psi_{\{\tilde I_i,\tilde J_i\}}\,,
\end{equation}
where 
\begin{equation}
\tilde I_i(K)=I_i(\tilde K),\quad \tilde J_i(K)=J_i(\tilde K)\,.
\end{equation}

\end{document}